\documentclass[12pt]{article}

\usepackage{aaspp4}
\usepackage[dvips]{graphicx}

\def\dfrac#1#2{{\displaystyle\frac{#1}{#2}}}

\makeatletter
\def\lnyoro{\mathrel{\mathpalette\gl@align<}}
\def\gnyoro{\mathrel{\mathpalette\gl@align>}}
\def\gl@align#1#2{\lower.6ex\vbox{\baselineskip\z@skip\lineskip\z@\ialign{$\m@th#1\hfil##\hfil$\crcr#2\crcr\sim\crcr}}}
\makeatother
\def\iris{{\sl IRIS}}
\def\iso{{\sl ISO}}
\def\cobe{{\sl COBE}}
\def\iras{{\sl IRAS}}
\def\dirbe{DIRBE}
\def\firas{FIRAS}
\def\lir{L_{\rm IR}}
\def\lirm{L_{\rm IR,\, max}}
\def\tlir{{\tilde{L}}_{\rm IR}}
\def\slim{S_{\rm lim}(\nu )}
\def\llim{L_{\rm IR, \, lim}}
\def\dl{d_{\rm L}}
\def\lf{\phi_0 (L_{\rm IR})}
\def\lfz{\phi (z, \,L_{\rm IR})}
 
\begin{document}

\title{The {\sl IRIS} Far-Infrared Galaxy Survey : \\
Expected Number Count, Redshift, and Perspective}

\author{\bf TSUTOMU T. TAKEUCHI$^{1,2}$, 
HIROYUKI HIRASHITA$^{1,2}$, 
KOUJI OHTA$^1$, 
TAKASHI G. HATTORI$^1$,
TAKAKO T. ISHII$^{1, 3}$,
}
\affil 
{$^1\;$  Department of Astronomy, Faculty of Science, Kyoto University,
Sakyo-ku, Kyoto 606--8502, JAPAN}
\affil
{$^2\;$  Research Fellow of the Japan Society for the Promotion of
Science}
\affil
{$^3\;$  Kwasan and Hida Observatories, Yamashina-ku, Kyoto University, 
Kyoto, \\607--8471, JAPAN}
\affil
{Electronic mail: takeuchi, hirasita, ohta, hattori, ishii@kusastro.kyoto-u.ac.jp}

\author{AND}

\author{\bf 
HIROSHI SHIBAI$^4$}
 
\affil
{$^4\;$  Division of Particle and Astrophysical Sciences, School of Science, 
Nagoya University, Chikusa-ku, Nagoya, 464--8602, JAPAN}
\affil
{Electronic mail: shibai@toyo.phys.nagoya-u.ac.jp}

\begin{abstract}

Infrared Imaging Surveyor (\iris) 
is a satellite which will be launched in the beginning of 2003.
One of the main purposes of the {\sl IRIS} mission is an all-sky survey 
at far-infrared (FIR) with a flux limit much deeper than that of {\sl IRAS}.
In order to examine the performance of the survey,
we estimated the FIR galaxy counts in four (50, 70, 120, and 150 $\mu$m) 
bands based on some models.
We adopted a multicomponent model which consists of
cirrus and starburst components for galaxy spectra, and the nearby 
FIR luminosity function derived from that of {\sl IRAS} galaxies.
We derived the number counts, redshift distributions, and infrared diffuse 
background radiation spectra for i) no evolution, ii) pure luminosity 
evolution, iii) pure density evolution with $
q_0 = 0.1$ and $0.5$.
We found that a large number of galaxies ( $\sim {\rm a\; few} \times 
10^6$ in the whole sky) will be detected in this survey.
With the aid of a vast number of detection, we will detect the effect of 
galaxy evolution, and evaluate the amplitude of evolution at least 
in the nearby universe in the \iris~survey, though it will be still 
difficult to constrain which type of evolution takes place from the 
number count alone.
We also studied the estimation of redshifts of detected galaxies by 
their infrared colors alone.
Although significant contamination takes place among nearby faint galaxies and
high-$z$ ones, we found that rough estimation of galaxy redshift can be 
practicable by jointly using present and future optical surveys.

\end{abstract}

\keywords{galaxies: evolution --- galaxies: formation --- galaxies: starburst 
--- infrared radiation}

\section{INTRODUCTION}

One of the important problems in astrophysics is to trace back the 
galaxy evolution to the epoch at which galaxies formed.
Recently Steidel and collaborators have pioneered and widely utilized
a multicolor technique to select young galaxies
by use of Lyman-break dropout, and discovered a class of actively
star-forming galaxies in high-redshift ($z \gnyoro 3$) Universe
(Steidel, Pettini,\& Hamilton 1995; Steidel et al. 1996; Lowenthal 
et al. 1997).
Some lens-magnified normal galaxies at high redshift also show rather 
high star formation rate (Ebbels et al. 1996; Yee et al. 1996; Franx 
et al. 1997; Trager et al. 1997).
Furthermore, Ohta et al. (1996) and Omont et al. (1996) detected
CO emission lines in the quasar BR1202--0725 at redshift $z = 4.69$,
suggesting the existence of a large amount of molecular gas and dust in 
the object.
All of these results imply that strong star formation and
synthesis of heavy elements have already occurred in such an early
stage of the Universe.
The violent initial bursts of star formation must have produced large
amount of dust, so that the optical deep surveys of protogalaxies
inevitably suffer from the bias induced by severe dust extinction.
The dust particles reprocess the starlight, and
bulk of the bolometric luminosity of these young galaxies will be
emitted at far-infrared (FIR) -- submillimeter (sub-mm) wavelengths.

The FIR emission from galaxies, especially from those are so faint that 
we are unable to resolve, produces the cosmic infrared background
radiation (CIRB). 
Therefore the CIRB provides important information on the past star 
formation history of galaxies
including the inaccessible sources (e.g. Bond, Carr, \& Hogan 1986;
Burigana et al. 1997).
The recently reported detection of an isotropic diffuse
light at $\lambda \sim 100 - 200 \mu$m has given this field new impetus 
(Puget et al. 1996; see also Fixsen et al. 1998; Hauser et al. 1998).

Meanwhile, galaxy evolution at low reshift ($z \lnyoro 1$) has also
crucial implications to cosmological studies.
Lilly et al. (1996) found the rapid rise of the comoving luminosity 
density with redshift in ultraviolet (UV), optical, and near-IR 
wavelenghts up to $z \sim 1$.
Madau et al. (1996) interpreted this as an increase of comoving star 
formation rate (SFR) by a factor of ten in the redshift interval 
$z = 0 - 1$.
Subsequent observational evidences support this conclusion (e.g. Ellis 
et al. 1996; Tresse \& Maddox 1998).
Because FIR is the wavelength at which dust reradiates the optical 
-- UV radiation from young stars, the steeply increasing SFR density
is expected to propose a significant increment of FIR source counts, 
which may be related to the evolving population of star-forming galaxies
discovered by {\sl Infrared Astronomical Satellite} (\iras$\,$) 
(Hacking \& Houck 1987 (HH87); Hacking, Condon, \&
Houck 1987; Saunders et al. 1990; Oliver,
Rowan-Robinson, \& Saunders 1992; Ashby et al. 1996).

For the understanding of FIR properties of actively
star-forming galaxies, vast advances have been provided by 
the \iras~all-sky survey 
with a flux limit of $\sim 1$ Jy at $60 \; \mu$m
(for a review, Soifer, Houck, \& Neugebauer 1987a).
Recently quite important results have been reported from the {\sl
Infrared Space Observatory} (\iso$\,$; Kessler et al. 1996).
Kawara et al. 1998 (K98) have performed a deep survey at the Lockman Hole 
region with a flux limit of 45 mJy at 175 $\mu$m.
They reported that the surface density of the sources 
brighter than 150 mJy at 175 $\mu$m approximately reproduced by the 
model of Guiderdoni et al. (1997).
Although this survey is much deeper than the \iras~survey, it covers only
a small solid angle of the sky (1600 arcmin$^2$) and much wider sky
coverage is necessary in the next step to obtain a huge sample.

Infrared Imaging Surveyor (\iris$\,$\footnote{The project name of \iris~is 
Astro-F.}) 
is a next generation infrared satellite, which will be launched in
the beginning of 2003.
A near-/mid-infrared camera (IRC) and a FIR scanner (FIS) are
planned to be on-board.
One of the main purposes of the \iris~mission is a FIR
all-sky survey deeper than \iras~surveys, and the FIR instrument represents
a significant improvement over \iras, so that much higher 
sensitivity is achieved (detection limit is $\sim 20$ mJy at $50\;\mu$m, 
$\sim 30$ mJy at $120\;\mu$m, and $\sim 50$ mJy at $150\;\mu$m).

In order to examine the performance of the survey,
we estimated the FIR galaxy counts based on a simple model with
various sets of cosmological parameters and evolution types.
There have been a number of attempts to predict the source
counts or the CIRB properties.
The models of the source counts and CIRB generally fall into two
categories (e.g. Eales \& Edmunds 1997):
one is tied closely to the nearby observational results but makes no
attempts to incorporate detailed galaxy evolution models
(e.g. Beichman \& Helou 1991 (BH91); Pearson \& Rowan-Robinson 1996).
These models are based on the local FIR luminosity function
(LF) and the observational evidence of spectral energy distribution
(SED) of galaxies in the FIR -- sub-mm, with the assumptions of simple
functional forms for the evolution.
The other is based on the models constructed by detailed physical
processes related to the evolution of galaxies (e.g. Franceschini et
al. 1994; Guiderdoni et al. 1997).
They tend to be very complex with large numbers of parameters, which
give rise to a number of related problems poorly understood.
Ellis (1997) names the former ``empirical approach'' and the latter
``{\sl ab initio} approach''.
Despite each category of models has its own merits and demerits, we
preferred the ``empirical approach'', because it is free of the 
parameters which are inherent in specific models and thus easy to 
understand the dependence on the cosmological and evolutional 
parameters.

This paper is organized as follows. In section 2, we describe the
general formulation of our model. 
The parameters to execute our calculations and their consistency 
with previous survey results are examined in section 3.
We present the {\sl IRIS}-based results and related discussions in
section 4, and our conclusions are summarized in section 5.

\section{MODEL DESCRIPTION}

\subsection{Spectral Energy Distribution of Galaxies}\label{sedg}

First we construct a model of galaxy SED from FIR to sub-mm wavelengths. 
The FIR emission originates from several dust components
of different grain sizes
and temperatures, reprocessing the interstellar radiation 
(D\'{e}sert, Boulanger, \& Puget 1990).
Rowan-Robinson \& Crawford (1989)(RC89) showed that the FIR spectrum 
of \iras~ galaxies is well fitted by the combination of two components: 
cool disk emission from interstellar dust (cirrus component) and hot 
starburst-induced emission, from optically thick dust clouds heated by 
numbers of  OB stars.
Therefore we consider a simple model of two components, cool cirrus and hot 
starbursts as the components of our model SEDs of galaxies.
We must note that we ignore the power-law component seen in AGN 
or QSO spectra, because contribution of these power-law objects is expected 
to be negligible in number compared with the total number of infrared (IR) 
sources, and because their luminosity contribution is also small in 
FIR wavelength range, though it might be slightly higher in very high-$z$ 
Universe (Pearson 1996).

We define the total IR luminosity $\lir$ as follows:
\begin{eqnarray}
L_{\rm IR}=\int_{\nu_{\rm min}}^{\nu_{\rm max}}L(\nu )\, {\rm d}\nu ,
\end{eqnarray}
where $L(\nu )$ is the monochromatic luminosity of a galaxy, 
and we put
$\nu_{\rm min}=3\times 10^{11}$ Hz ($\lambda_{\rm max} = 1$ mm) 
and $\nu_{\rm max}=10^{14}$ Hz ($\lambda_{\rm min} = 3 \; \mu$m).

Here we assume that the IR luminosity of a galaxy less than
$10^{10}L_\odot$ is composed of the cirrus component $L_{\rm c}(\nu )$
only. 
For galaxies with $L_{\rm IR} > 10^{10}L_\odot$, 
the IR luminosity in excess of $10^{10} \; L_\odot$
(i.e., $L_{\rm IR}-10^{10}L_\odot$), 
is assumed to come from the starburst component, denoted as 
$L_{\rm s}(\nu )$. 
Thus, the composite SED of a galaxy is expressed as follows;
\begin{eqnarray}
L(\nu )=\left\{
\begin{array}{ll}
L_{\rm c}(\nu ) & \mbox{for} \ L_{\rm IR}<10^{10}L_\odot ;\\
L_{\rm c}(\nu )+L_{\rm s}(\nu ) & \mbox{for} \ L_{\rm
IR}>10^{10}L_\odot ,
\end{array}
\right.
\end{eqnarray}
where $L_{\rm c}(\nu )$ and $L_{\rm s}(\nu )$ are normalized as
\begin{eqnarray}\label{cirrusspec}
L_{\rm c}\equiv\int_{\nu_{\rm min}}^{\nu_{\rm max}}L_{\rm c}(\nu )\,
{\rm d}\nu =\left\{
\begin{array}{ll}
L_{\rm IR} & \mbox{for} \ L_{\rm IR}<10^{10}L_\odot ; \\
10^{10}L_\odot & \mbox{for} \ L_{\rm IR}>10^{10}L_\odot ,
\end{array}
\right.
\end{eqnarray}
and
\begin{eqnarray}
L_{\rm s}\equiv\int_{\nu_{\rm min}}^{\nu_{\rm max}}L_{\rm s}(\nu )\,
{\rm d}\nu =\left\{
\begin{array}{ll}
0 & \mbox{for} \ L_{\rm IR}<10^{10}L_\odot ; \\
L_{\rm IR}-10^{10}L_\odot & \mbox{for} \ L_{\rm IR}>10^{10}L_\odot .
\end{array}
\right.
\end{eqnarray}
This model formulation is in line of BH91.
So as to make our model work well, 
$\lir$ should be uniquely defined from the value $L(\nu)$ at any 
frequency $\nu$.
The SEDs depend only on $\lir$, and do not vary with redshift as long as
$\lir$ of a galaxy is constant.

\subsubsection{The cirrus component}

The cirrus spectrum we used is based on the model of D\'{e}sert et al. 
(1990).
Their model spectrum is the sum of three subcomponents, that are 
polycyclic aromatic hydrocarbons (PAHs; peaked at $\sim 10 \; \mu$m), 
very small grains (peaked at $\sim 60 \; \mu$m), and big grains 
(peaked at $\sim 100\; \mu$m).
No IR emission lines are included in our model.
As mentioned above, galaxies with
$L_{\rm IR}<10^{10}L_\odot$ only have the cirrus component in the present 
model.
At each IR luminosity, the galaxy SED is scaled properly to yield the given 
$\lir$ according to eq.(\ref{cirrusspec}).

\subsubsection{The Starburst Component}

As for galaxies with $L_{\rm IR} \gnyoro 10^{10 - 11}L_\odot$, 
bulk of the IR luminosity is due to dust emission from an intense starburst
in giant molecular clouds (e.g. Sanders \& Mirabel 1996).
Therefore we assume that, for galaxies whose IR luminosities are 
brighter than $10^{10} L_\odot$, their IR luminosities 
in excess of $10^{10}L_\odot$ consist of starbursts with two 
different temperatures according to BH91.
We give the starburst flux $L_{\rm s}(\nu )$ as a superposition of thermal
blackbody radiation spectrum of temperature $T_{\rm hot}$ (we call it ``hot 
component'' of starbursts) and that of  $T_{\rm cool}$ (``cool component''), 
multiplied with dust emissivity.
We adopted the wavelength-dependent dust emissivity $\propto \lambda^{-1} 
\propto \nu $ (RC89).
By using the above temperatures, $L_{\rm s}(\nu )$
is expressed by 
\begin{eqnarray}
L_{\rm s}(\nu )=\alpha\nu B_\nu (T_{\rm cool})+\beta\nu B_\nu
(T_{\rm hot}),
\end{eqnarray}
where $B_\nu (T)$ is the Planck function with temperature $T$ and both 
$\alpha$ and $\beta$ are normalizing constants (RC89). 
The temperatures of the starburst components are given by
\begin{eqnarray}
T_{\rm cool}=60\left(\frac{L_{\rm s}}{10^{11}L_\odot}\right)^{0.1}\,
{\rm K},
\end{eqnarray}
and
\begin{eqnarray}
T_{\rm hot}=175\left(\frac{L_{\rm s}}{10^{11}L_\odot}\right)^{0.1}\,
{\rm K}
\end{eqnarray}
(BH91; see also Rieke \& Lebofsky 1986; Helou 1986;
Soifer et al. 1987b). 
The cool component gives a good representation
of the larger (0.01--0.1 $\mu$m) grains modeled by Rowan-Robinson
(1986), and the hot component gives an approximate representation
of the very small grains postulated by Boulanger, Baud, \& van Albada 
(1985), Draine \& Anderson (1985), and Rowan-Robinson (1992).

Normalizing constants, $\alpha$ and $\beta$, 
are determined by the following equations (BH91):
\begin{eqnarray}
0.7L_{\rm s} & = & \alpha\int_{\nu_{\rm min}}^{\nu_{\rm max}}\nu B_\nu
(T_{\rm cool})\,{\rm d}\nu , \\
0.3L_{\rm s} & = & \beta\int_{\nu_{\rm min}}^{\nu_{\rm max}}\nu B_\nu
(T_{\rm hot})\,{\rm d}\nu .
\end{eqnarray}

The model SEDs of galaxies with $\lir = 10^8 L_\odot - 10^{14} 
L_\odot$ constructed in this manner are shown in Fig. 1. 
The solid curves represent our model SEDs with various $\lir$ (the lowest one:
$\lir = 10^8 L_\odot$; level interval: $\Delta \log \lir = 1.0$).
We also presented the observed SEDs of three representative IR galaxies, 
M82 (Klein, Wielebinski, \& Morsi 1988), IRAS F$10214+4724$ 
($z=2.286$, Rowan-Robinson et al. 1993), and SMM$02399-0136$ 
($z=2.803$, Ivison et al. 1998).
Open triangles stand for the observed SED of M82,
open squares depict the SED of IRAS F$10214+4724$, and 
open diamonds represent the SED of SMM$02399-0136$.
Small downward arrows show upper-limit values.
The IR luminosity of M82 is $\sim 4 \times 10^{10} L_\odot$ and that of 
IRAS F$10214+4724$ is $\sim 3 \times 10^{14} L_\odot$, but the latter has 
been revealed to be magnified by gravitational lensing by a factor of 
$\sim 20$ (Broadhurst \& Leh\'{a}r 1995).
Its $\lir$ turns out to be $\sim 10^{13} L_\odot$, and its SED is corrected 
in Fig. 1.
SMM$02399-0136$ is also a lensed object, and its unlensed $\lir$ is 
$\sim 4 \times 10^{13}$ (Ivison et al. 1998).
The SED of this galaxy is also corrected for the lensing (by factors of 3.5).
Their SEDs are approximately fitted by our models with corresponding $\lir$.

\subsection{Local Luminosity Function}

We adopted the LF based on the \iras~ by Soifer et al. (1987b)
as the local IR LF of galaxies . 
Their LF is defined by $L_{60\, \mu {\rm m}}$, so we converted it into our 
$\lir$, using our model SEDs.
Then we performed an analytical fitting to the resultant data points,
and obtained the following double power-law form for our LF:
\begin{eqnarray}
\log [h_{75}^3 \phi_0(h_{75}^{-2} L_{\rm IR})]=\left\{
\begin{array}{ll}
7.9-1.0\log\left({h_{75}^{-2}L_{\rm IR}}/{L_\odot}\right) & \mbox{for} \
10^8 L_\odot < h_{75}^{-2} L_{\rm IR} < 10^{10.3}L_\odot ; \\
17.1-1.9\log\left({h_{75}^{-2} L_{\rm IR}}/{L_\odot}\right) & \mbox{for} \
10^{10.3}L_\odot < h_{75}^{-2} L_{\rm IR} < 10^{14}L_\odot ; \\
{\rm no \; galaxies}  &  \mbox{otherwise},\label{lf}
\end{array}
\right.
\end{eqnarray}
where $\phi_0$ is the number density of galaxies in Mpc$^{-3}$
dex$^{-1}$ in case of $h_{75} = 1$, where $h_{75}$ is the Hubble 
parameter normalized by 75 km s$^{-1}$ Mpc$^{-1}$ (i.e. $h_{75}=H_0/75$ km
s$^{-1}$ Mpc$^{-1}$).
This LF is presented in Fig. 2. 
We note that, because the volume element (defined in eq.\ref{kolb}) depends 
on $H_0^{-3}$, luminosity function multiplied by the 
volume element has no dependence on the Hubble parameter. 
This means that neither galaxy number count nor the expected redshift 
distribution depend on the Hubble parameter in the empirical approach. 
Thus, we do not have to consider the effect of the Hubble parameter
in the following analyses.

The faint-end slope of the LF of Soifer et al. (1987b) is much steeper than 
that derived by Saunders et al. (1990).
Little is known about the population which occupies the faint end of the LF.
We discuss the effect of it in sections \ref{zd} and \ref{ccdiag}.

\subsection{Treatment of Galaxy Evolution}\label{evpar}

For galaxy evolution, we deal with the following three cases:
(i) no evolution (ii) pure luminosity evolution, and (iii) pure 
density evolution.
We define the LF at $z$, $\lfz$ in {\sl comoving} volume.
Then various types of galaxy evolution are characterized by the form of 
the functions which represent the relation between $\lf$ and $\lfz$.
We also define cumulative LF at redshift $z$, $\Phi (z\,,\;\lir)$,
\begin{eqnarray}
\Phi(z\,,\;\lir) &=& \int_{L_{\rm IR}}^\infty\phi (z,\, \lir')\,
{\rm d}\log \lir'\; \\ \label{clf}
 &=& \int_{L_{\rm IR}}^{\lirm} \phi (z,\, \lir')\,
{\rm d}\log \lir'\;, \nonumber
\end{eqnarray}
where $\lirm$ is the maximum IR luminosity at the epoch.

\subsubsection{No evolution}

The term ``no evolution'' means that the luminosity function is
independent of $z$:
\begin{eqnarray}
\phi (z,\, L_{\rm IR})=\phi_0(L_{\rm IR}).
\end{eqnarray}
Thus the cumulative LF reduces to 
\begin{eqnarray}
\Phi(z\,,\;\lir) = \int_{\lir}^\infty\phi_0(\lir')\,
{\rm d}\log \lir'\;.
\end{eqnarray}

Although no evolution model seems to describe an unphysical 
situation, its prediction acts as a 
valuable standard baseline from which the various evolutionary 
differences can be compared.

\subsubsection{Pure luminosity evolution}

The effect of the luminosity evolution of galaxies is modeled by
\begin{eqnarray}
  &\lir(z) = \lir(0)f(z)&, \label{deflev}\\ 
&f(z)\equiv\exp\left[ Q\dfrac{\tau (z)}{t_{\rm H}}\right]\; ,&
\end{eqnarray}
where $Q$, $\tau (z)$ and $t_{\rm H}$ are a parameter defining the
magnitude of evolution,
look-back time as a function of $z$, and the Hubble time $1/H_0$.
This functional form of evolution is proposed by Broadhurst, Ellis, 
\& Glazebrook (1992).
Peacock (1987) derived a useful approximation for the look-back time:
\begin{eqnarray}\label{lbt}
\frac{\tau (z)}{t_{\rm H}}\simeq\frac{1}{\beta}\left(1-(1+z)^{-\beta}
\right),
\end{eqnarray}
where
\begin{eqnarray}
\beta = 1 + \frac{(2q_0)^{0.6}}{2}.
\end{eqnarray}
This expression for the look-back time is 
exact for $q_0 =0$ and $q_0
=1/2$ and is accurate within the error of $\lnyoro 1$ \% for
$q_0\lnyoro 3/2$ (Peacock 1987).
The function reduces to the simpler form $f(z)\simeq (1+z)^Q$ 
at the low-$z$ epoch.

Pure luminosity evolution leads to the following expressions as 
cumulative LF,
\begin{eqnarray}
\Phi(z\,,\;\lir) &=& \int_{\lir}^{\infty}\phi \left(z,\lir' \right)\,
{\rm d}\log \lir' \nonumber\\ 
 &=& \int_{\lir/f(z)}^{\infty} \phi_0 (\tlir')\,
{\rm d}\log \tlir'\;, \label{lclf} \\
\tlir &=& \dfrac{\lir}{f(z)}\; . \label{lirf}
\end{eqnarray}

\subsubsection{Pure density evolution}

We, here, consider the model that only the comoving number
density of galaxies changes as a function of redshift, 
which is called ``pure density evolution''.
Pure density evolution is expressed as
\begin{eqnarray}
\phi(z, \lir) = \phi_0(\lir)g(z)\; .\label{defdev}
\end{eqnarray}
The function $g(z)$ is defined by
\begin{eqnarray}
g(z) \equiv \exp \left[P\frac{\tau (z)}{t_{\rm H}}\right],
\end{eqnarray}
where $P$ is a parameter defining the amplitude of evolution.
This is the same type of function considered in the previous subsection.
We note that this type of galaxy evolution is not aimed at describing
the ``merging'' process.
The number count model including merging process is widely discussed  
by Gardner (1998).
The cumulative LF becomes
\begin{eqnarray}
\Phi(z\,,\;\lir) = \int_{\lir}^\infty \phi_0 (\lir') g(z)\,
{\rm d}\log \lir'\;. \label{dclf}
\end{eqnarray}

\subsection{Galaxy Number Count}

Using the above formulae, now we calculate the 
flux--number relation, or so-called galaxy number count.
We assume that galaxies are regarded as point sources (i.e. 
cosmological dimming of surface brightness is not taken into account).
Then the relation between observed flux $S(\nu)$ and emitted monochromatic 
luminosity $L(\nu)$ is given by 
\begin{eqnarray}
S(\nu ) =\frac{(1+z)L\left( (1+z)\nu\right)}{4\pi d_{\rm L}^2}\,,\label{detec}
\end{eqnarray}
where $d_{\rm L}$ is luminosity distance.
In the Universe without the cosmological constant $\Lambda$, 
$\dl$ is expressed by analytical form, in terms
of redshift $z$, deceleration
parameter $q_0$ and Hubble parameter $H_0$ (Mattig 1958);
\begin{eqnarray}
d_{\rm L}=\frac{c}{H_0q_0^2}[zq_0+(q_0-1)(\sqrt{2q_0z+1}-1)]\; ,
\end{eqnarray}
($c$ : the velocity of light).
When we fix a certain $S(\nu)$, we obtain $L((1 + z) \nu)$ by using 
eq. (\ref{detec}).
Then the correspondent $\lir(S(\nu),\; z)$ at the redshift $z$ is 
uniquely determined (see \ref{sedg}).

We define $N(> S (\nu))$ as the number of galaxies with a detected 
flux density larger than $S (\nu)$, then it is formulated as
\begin{eqnarray}
  N(>S(\nu)) = \int_\Omega {\rm d}\Omega\int_0^{z_{\rm max}}{\rm d}z
  \frac{{\rm d}^2V}{{\rm d}z\,{\rm d}\Omega}
  \int_{L_{\rm IR}(S(\nu)\,, z)}^{\infty}
  \phi(z,\, \lir') \; {\rm d}\log \lir'\; , \label{gnc}
\end{eqnarray}
where ${{\rm d}^2V}/{{\rm d}z}{\rm d}\Omega$ is the {\sl comoving} 
volume element per str per $z$, which can be expressed in terms of 
cosmological parameters as 
\begin{eqnarray}
  \left.\frac{{\rm d}^2V}{{\rm d}\Omega\,{\rm d}z}\right|_z=
  \frac{c}{H_0}\frac{d_{\rm L}^2}{(1+z)^3\sqrt{1+2q_0z}},
\label{kolb}
\end{eqnarray}
(e.g. Kolb \& Turner 1994).
We set $z_{\rm max}$ as 10.

Next, we formulate the number--flux relation
with our evolution models.
The no-evolution prediction is as follows:
\begin{eqnarray}
  N(> S (\nu)) = \int_\Omega {\rm d}\Omega\int_0^{z_{\rm\,max}}{\rm d}z
  \frac{{\rm d}^2V}{{\rm d}z\,{\rm d}\Omega} 
  \int_{\lir (S (\nu)\,, z)}^{\infty}
  \phi_0(\lir')\; {\rm d}\log \lir'\; .
\end{eqnarray}
If pure luminosity evolution takes place, the expected number 
count is derived from eqs.(\ref{lclf}) and (\ref{gnc}) ,
\begin{eqnarray}
  N(> S (\nu) ) = \int_\Omega {\rm d} \Omega \int_0^{z_{\rm max}}{\rm d}z
  \frac{{\rm d}^2V}{{\rm d}z\,{\rm d}\Omega}
  \int_{\lir(S(\nu), z)/f(z)}^{\infty} \;
  \phi_0 (\tlir')\; {\rm d} \log \tlir'\; .
\end{eqnarray}
The expression for pure density evolution is obtained in the same 
manner;
\begin{eqnarray}
  N(> S (\nu) ) = \int_\Omega {\rm d} \Omega \int_0^{z_{\rm max}}{\rm d}z
  \frac{{\rm d}^2V}{{\rm d}z\,{\rm d}\Omega}
  \int_{\lir(S(\nu), z)}^{\infty} \phi_0(\lir')g(z) \;
  {\rm d} \log \lir'\; .
\end{eqnarray}

\subsection{Redshift Distribution}\label{zd}

When the flux detection limit $\slim$ is given, the limiting luminosity 
at redshift $z$ can be obtained by the following process.
From equation (\ref{detec}), the limiting monochromatic luminosity 
$L_{\rm lim}(\nu_{\rm em})$, where $\nu_{\rm em}$ is the emitted 
frequency at the
rest frame of a galaxy at redshift $z$, can be calculated as
\begin{eqnarray}
L_{\rm lim}(\nu_{\rm em})=\frac{4\pi d_{\rm L}^2}{1+z} S_{\rm lim}
\left(\frac{\nu_{\rm em}}{1+z}\right).
\end{eqnarray}
We, then, obtain the total IR luminosity $L_{\rm IR,\, lim}$, which 
uniquely corresponds to the given $L_{\rm lim}(\nu_{\rm em})$, 
and galaxies with $L_{\rm IR}$ larger than $L_{\rm IR,\, lim}$ are
detected.
In terms of the derived $\llim$, the redshift distribution is 
formulated as
\begin{eqnarray}
\left.\frac{{\rm d}^2N}{{\rm d}\Omega\,{\rm d}z}\right|_z 
&=&\left.\frac{{\rm d}^2V}{{\rm d}\Omega\,{\rm d}z}\right|_z
\int_{\llim}^\infty \phi (z, \lir')\, {\rm d}\log \lir' \nonumber \\ 
&=&
\left.\frac{{\rm d}^2V}{{\rm d}\Omega\,{\rm d}z}\right|_z
\Phi(z, \llim) \label{zdist}.
\end{eqnarray}
The evolutionary effect can be easily introduced in 
the redshift distribution by using eqs. (\ref{deflev}) or  
(\ref{defdev}), as $\Phi (z, \lir)$ in equation 
(\ref{zdist}), and substituting 
$\llim$ instead of  $\lir$.

\section{PARAMETERS FOR THE EVOLUTIONARY MODELS}

In this section, we estimate the evolutionary 
parameters $P$ and $Q$ defined in section \ref{evpar} in order to 
perform our calculation.
Then we check the validity and consistency of our predictions by comparison 
with the \iras- and \iso-based number counts and the CIRB spectrum
obtained by \cobe~.

\subsection{Evolutionary Parameters}

We used \iras~extragalactic source count data to obtain the values of 
parameters $P$ and $Q$.
Among \iras~bandpasses (12, 25, 60, and 100 $\mu$m), the 60-$\mu$m band
is known to be the most suitable for extragalactic studies, and 
intensively used in the context of the investigation of galaxies 
(e.g. Soifer et al. 1987a).
Therefore we employed the 60-$\mu$m data of \iras~galaxies.

We first present the differential flux-number
relation of \iras~60-$\mu$m databases, \iras~Point Source Catalog (\iras~PSC)
and HH87.
This is depicted in Fig. 3.
Open squares represent the data points of \iras~PSC, and filled squares,
HH87.
These diagrams are constant against flux if the Universe is static Euclidean
and source distribution is homogeneous.
The prediction of no-evolution model is also shown in Fig. 3.
We perform least-square fitting of our evolutionary models on
these data points.
When we use all data points with equal weight, the most likely values 
for the parameters are $(P\,,\; Q) = (1.3\,,\; 0.7) \; (q_0 = 0.1)$ and 
$(1.4\,, \; 0.8) \; (q_0 = 0.5)$.
We obtain higher values $(P\,,\; Q) = (2.7\,,\; 1.4) \; (q_0 = 0.1)$ and 
$(2.8\,, \; 1.5) \; (q_0 = 0.5)$ when the four statistically poorest points
around $\log S\; [{\rm Jy}] \sim -0.5$ are omitted.
We adopt the latter for our studies on evolution.
We note that the estimation of $P$ and $Q$ depends on the fiducial point 
which is used to determine the normalization of galaxy counts.
Some authors define the parameters $P$ and $Q$ so that the model 
predictions have the same value of $({\rm d}N/{\rm d}S) \, S^{2.5}$ 
at 1 Jy (see e.g., Fig. 1 of Oliver et al. 1992). 
The others estimated $P$ and $Q$ by overall fitting, as the same 
as we did, and did not made such kind of normalization (see e.g., Fig. 5 of
Ashby et al. 1996).
Therefore our values for $P$ and $Q$ 
seem smaller than those who adopted the former method, but the effect 
of the evolution is almost the same.
These parameters are summarized in Table 1.
They are almost independent of $q_0$, at least within the available flux 
limit of \iras~data.

\subsection{Galaxy Count : Comparison with \iras~and \iso~Results}

We compared our number count predictions
with the previous survey results obtained by \iras~and \iso.
In the upper two panels of Fig. 4, the 60-$\mu$m number 
count predictions 
with two evolutionary models $(q_0 = 0.1 \; {\rm and} \; 0.5)$ are 
presented with the result of 
QMW \iras~galaxy survey (Rowan-Robinson et al. 1991).
Within the depth of the QMW catalog there is only a small difference between 
no-evolution and evolutionary predictions, and the model predictions agree 
with the observed counts.
The evolutionary effect is significant in fainter flux regime (e.g. Bertin, 
Dennefeld, \& Moshir 1997).

Comparison between our 175-$\mu$m galaxy number count and the results
of the \iso~Lockman Hole survey (K98) is illustrated in the lower panels of 
Fig. 4.
Source identification and flux calibration contains considerable 
difficulties, which turned out to be an uncertainty of the number count
by a few factors.
Both \iras~and \iso~number counts are successfully reproduced by our 
model calculations.

\subsection{Cosmic Infrared Background}\label{coirb}

The CIRB is generated from the integrated light of galaxies.
Therefore, combining the SEDs of galaxies and number count predictions, 
we obtain the CIRB spectrum.
The observed flux density of a galaxy whose IR luminosity is $\lir$, 
$S(\nu, \lir)$ is given by equation (\ref{detec}) as follows:
\begin{eqnarray}
S(\nu \,,\lir)=\dfrac{(1 + z)L\left(\nu (1+z)\,, \lir \right) }
{4\pi d_{\rm L}^2},
\end{eqnarray}
where $L(\nu , \lir)$ is the monochromatic luminosity of a galaxy with 
$\lir$.
Then the CIRB spectrum $I(\nu)$, i.e., the background flux density from 
unit solid angle, is expressed as
\begin{eqnarray}
  I(\nu )=\int_0^{z_{\rm max}} {\rm d}z 
  \frac{{\rm d}^2V}{{\rm d}z\,{\rm d}\Omega}
  \int_0^\infty\phi (z,\, \lir') S(\nu\,, \;\lir') \; 
  {\rm d}\log \lir' \;.
\end{eqnarray}

We can deal with the evolutionary effect on $I(\nu)$ through $\lfz$, just
the same as in the case of number count calculation.
The expected CIRB spectra are presented in
Fig. 5.
The upper panel shows the case of $q_0 = 0.1$, and lower panel, $q_0 = 0.5$.
We also put the observational constraints on $I(\nu)$ obtained by \cobe~
measurement in Fig. 5.
Red open triangles represent the ``dark sky'' upper limit to the CIRB 
measured by \cobe~Diffuse Infrared Background Experiment (DIRBE).
The \dirbe~sky brightness varies roughly sinusoidally over the year, 
due to the complex features of the interplanetary dust cloud.
The ``dark sky'' brightness is that of the darkest area on the sky at each
wavelength.
Filled squares are the result of a similar analysis of the \cobe~Far 
Infrared Absolute Spectrophotometer (\firas) high-frequency data, 
after removal of the CMB (cosmic microwave background) signal.
The dark sky values of \dirbe~and \firas~show excellent agreement with 
each other.
Blue open triangles represent the residual signal of \dirbe~after removing 
the contributions from the model foreground sources such as Galactic diffuse 
emission, and interplanetary dust emission.
As described by Mather et al.(1994), the CMB spectrum in the wavelength 
range $0.5 - 5$ mm deviates from a 2.726 K-blackbody shape by less than 
0.03 \% of the peak intensity.
This deviation is shown by green horizontal lines.
Details of these background signals are summarized in Hauser (1995).
In addition, Puget et al. (1996) have reported the isotropic diffuse
residual light of \firas~data, after subtracting the foreground 
contributions. 
Fixsen et al. (1998) and Hauser et al. (1998) have also reported 
similar results.
The orange broken solid lines are the approximate shape of the 
spectrum of the residual light reported by Puget et al., Fixsen et al., 
and Hauser et al. with errors.

Within the wavelengths shorter than 300 $\mu$m, our model prediction 
is consistent with the CIRB intensity, but in the longer wavelengths, 
it seems to violate the limits from the CIRB (larger $q_0$ mitigates 
this discrepancy).
In such a long wavelength regime, the background radiation is dominated by 
the contributions from hyperluminous IR galaxies at very high redshifts.
As we see from Fig. 1, the SEDs of such luminous galaxies are well 
represented by our SED models.
Thus, the functional form of the galaxy evolution might be too simple 
and possibly not valid in extremely high-$z$ objects, or
alternatively, high-luminosity end of the LF might not be valid at high-$z$. 
In spite of the difficulty in sub-mm range, our model works for our 
present purpose which is IR source number count prediction in wavelength
region shorter than 300 $\mu$m.

\section{RESULTS AND DISCUSSIONS}

\subsection{{\sl IRIS} Survey Performance}

\subsubsection{Flux Detection Limit of \iris~FIS}\label{fislim}

The flux detection limit $\slim$ is decided by some factors;
the sensitivity of the device, the intervening emission, e.g.
zodiacal light, interstellar emission of Our Galaxy, and source 
confusion etc.
The dominant factor is different in different wavebands, and the resultant
$\slim$ is dependent on wavelength.
Thus the calculation requires the careful consideration of $\slim$
individually for each $\lambda$ or $\nu$.
We assume 50-, 70-, 120-, and 150$\mu$m bandpasses for \iris~FIS.
Unstressed gallium-doped germanium (Ge:Ga) is used for the shorter 
wavelength in FIR $(50 - 110\,\mu{\rm m})$ and stressed Ge:Ga for the longer
wavelength $(\lnyoro 170\,\mu{\rm m})$ (Kawada et al. 1998).
In the shorter wavelength, sensitivity is limited by internal and background
noises, while in the longer wavelength, it is constrained by the source and
Galactic cirrus confusion (Gautier et al. 1992; Thronson et al. 1995).
The resultant $5\sigma$-detection limits of the assumed four wavebands are 
summarized in Table 2\footnote{The flux detection limits presented here are
based on those in Kawada et al. (1998), but are slightly (by a factor of 1.5 
-- 2) deeper than the limits they reported; the values adopted here are 
those expected in the high Galactic latitude region ($|b| > 60^\circ$).
Much better point source detection limits 
can be expected in limited sky areas near the ecliptic poles, where the survey 
scan will be repeated more than hundred times, and can be achieved by a 
spatial deconvolution algorithm.}.

\subsubsection{\iris~galaxy number count predictions}

Our predictions for the galaxy number count at assumed \iris~four 
bandpasses are shown in Fig. 6.
The upper four panels depict the number counts with $q_0 = 0.1$ and the 
lower four, $q_0 = 0.5$; 
from left to right, $\lambda = 50, 70, 120,$ and $150\,\mu$m.
The black solid lines show the no-evolution prediction.
The red and cyan lines represent the case of pure luminosity 
evolution and pure density evolution, respectively.
The green dot-dashed lines in Fig. 6 denote the \iris~FIS flux 
detection limit at each band.
The integrated detection number to the flux limit of each bandpass is 
$\sim 10^5 \; {\rm str^{-1}}$ at $50 \; \mu$m, $\sim 2 \times 10^5 \; 
{\rm str^{-1}}$ at $70 \; \mu$m, $\sim 3 \times 10^5 \; {\rm str^{-1}}$ 
at $120 \; \mu$m, and $\sim 2 \times 10^5 \; {\rm str^{-1}}$ at $150 \; 
\mu$m in the case of no evolution, 
thus the expected detection number is largest at $120\;\mu$m, and the total 
number of galaxy is expected to be $\sim {\rm several} \times 10^6$ in the 
whole sky.
This is about 100-times larger than the number of \iras~galaxies, which is 
$\sim 25000$ in all the sky.
It is clearly seen that 
$q_0$ does not affect the number count within the detectable flux range 
of \iris~ as shown in Fig. 6.

The effect of evolution appears most significantly at $70$ $\mu$m among 
the assumed bands, and the total detection number becomes twice larger.
Thus, the determination of the amplitude of galaxy evolution is best 
estimated at around $70\;\mu$m in the wavelength and flux coverage of 
\iris.
With the aid of such a vast number of galaxy detections, we will be able to 
evaluate the evolution strength ($P$ or $Q$) precisely.
In the case of $70$-$\mu$m band, statistical fluctuation
becomes much smaller than those of previous area-limited deep surveys.
Consequently the errors in the estimation of $P$ or $Q$ will be small enough, 
at least, to determine them in the relatively low-redshift ($z \lnyoro 1$) 
universe.
Unfortunately, 
there exists no drastic difference between 
luminosity and density evolution in the behaviors of number count,
and judging which evolution takes place from number count alone 
still remains as a difficult work.

\subsection{Redshift Distribution}\label{reddist}

Using the \iris~FIS flux detection limits we gave in section \ref{fislim}, 
we calculate the redshift distribution of galaxies detected by the 
\iris~all-sky survey.
Figure 7 shows the expected redshift distribution of galaxies 
detected in the assumed four bands.
Figure 7a illustrates the prediction at 50-$\mu$m band.
Figures 7b, 7c, and 7d are the same as Fig. 7a, except that they show 
the predictions at 70-, 120-, and 150-$\mu$m bands, respectively.
Solid curves denote the predictions for $q_0 = 0.1$, and dashed curves, 
$q_0 = 0.5$.
Larger $q_0$ leads to deeper $z$-distribution.
In any bandpasses, the numbers of galaxies at $z \gnyoro 1$ are $\sim 
{\rm a\; few} \times 10^3$, and several times larger with evolution.
Namely, low-$z$ galaxies are the overwhelming majority among the detected 
objects.
The steep increases of galaxy number at $z \lnyoro 0.5$ seen in 
Figs. 7c and 7d are attributed to the peak at $\sim 100\; \mu$m, which 
originates from big grains, in the SED of low-$\lir$ ($\lir \lnyoro 
10^{11}\; L_\odot$) galaxies. 
Figure 8 is an example of the redshift distribution for detected galaxies 
with various $\lir$ ($q_0 = 0.1$, no evolution).
Such galaxies can be detected only within $z \lnyoro 0.5$.
The LF we employed makes them to be abundant; if we use flatter faint-end 
slope of the LF, the increases would be less prominent.

The overall shape of the redshift distribution depends on the observing 
bandpass wavelength.
Nearest galaxies ($z \sim 0 - 0.5$) are more likely to be 
detected in longer wavelength $\lambda = 120 - 150\;\mu$m, and their 
detection number steeply decreases with increasing $z$ in the longest
wavelengths.
Galaxies which lie within the redshift range of $z \sim 0.5 - 2.5$ are 
expected to be detected more at 70 $\mu$m than at the other bands.
When the galaxy redshift exceeds $\sim 3$, they become harder to be detected
in the shorter bands, while they remain detectable in the longer bands.
The detection numbers at $\lambda = 120\;\mu$m and $150\; \mu$m, therefore, 
surpass those at $\lambda = 50\;\mu$m and 70 $\mu$m for the furthest galaxies.
This is understood as follows.
At significant redshift ($z \gnyoro 3$), the detected objects are the most 
luminous starbursts, whose SEDs peak at around $30\;\mu$m in the rest frame.
Since the peaks of luminous starbursts shift away from their rest 
wavelength, they are not detected at high-$z$ in shorter wavelengths.
Instead, in the longer wavebands, redshift effect makes their observed flux 
density nearly constant (known as the negative $K$-correction), and they 
can be still seen at high-$z$.

\subsection{Color--color Diagram in FIR} \label{ccdiag}

As we saw in section \ref{reddist}, most of the objects detected in 
the \iris~galaxy survey are nearby ones.
Hence, when we focus on the very young or primeval galaxies, an 
efficient method to pick up such candidates without knowing 
their spectroscopic redshifts will be very useful.
We, here, make attempts to utilize the peaks of the SEDs to 
estimate the redshifts roughly from their FIR colors alone.
We suppose that all the Galactic objects are taken away by certain 
methods.
This may well be possible at high Galactic latitudes.

Figure 9 shows the loci of galaxies with various $\lir$ on the FIR 
color--color ($\log S_{50}/S_{70}$ -- $\log S_{50}/S_{150}$) plane, where 
$S_\lambda$ is the detected flux density of galaxies at wavelength 
$\lambda$ [$\mu$m] in unit of $[{\rm erg\,s^{-1}cm^{-2}Hz^{-1}}]$.
Filled symbols are put on every $\Delta z = 0.5$ interval from $z = 0$.
When the sources fade away from the \iris~detection limits, we do not 
show their loci and symbols.
Red filled squares represent the colors of the low-$z$ ($z < 1$) galaxies, 
green filled squares the intermediate-$z$ ($1 < z < 2.5$), 
and blue filled circles the high-$z$ ($2.5 < z$) ones.
Because of the way of the construction of our model SEDs, 
galaxies with $\lir = 10^8 - 10^{10}\, L_\odot$ have the same colors
(the red filled square at the top-right).
To see how the existence of starburst component changes the color of cool 
cirrus galaxies, we put the galaxies of $\lir = 10^{10.1}\,L_\odot - 
10^{10.9}\, L_\odot$ with an interval of $\Delta \log \lir = 0.1$ 
(denoted by open squares from the top-right to down-left).
They correspond to the galaxies whose IR luminosity fraction of the 
superposed starburst is 20.6, 36.7, 50.0, 60.2, 68.4, 74.9, 80.0, 84.2, and 
87.4 \%, respectively.

When we consider the color selection criteria, the intrinsic scatter of 
galaxy color has to be taken into account.
Schmitt et al. (1997) proposed the template SEDs of various classes 
of galaxies.
Their SEDs show that the dispersion at FIR wavelength is an order 
unity.
The scatter of \iras~$S_{100}/S_{60}$ is also known to be an order of 
magnitude (RC89).
Considering these scatters, we can estimate the domains where galaxies 
locate on the FIR color-color plane.
We show the the domains of low- ($z < 1$), intermediate- ($1 < z < 2.5$) 
and high-$z$ ($z > 2.5$) galaxies with a red dashed rhomboid, a green 
solid rectangle, and a blue solid rectangle, respectively.
The high-$z$ and mid-$z$ regions are well separated on this diagram.

We must think of the overlap between low-$z$ and mid/high-$z$ domains.
The low-$z$ population creeping in the mid/high-$z$ domain consists of 
only low-$\lir$ galaxies ($\lir \lnyoro 10^{11}\,L_\odot$).
As we have seen in section \ref{reddist}, 
such galaxies quickly become undetectable with increasing redshift and 
reside in $z \lnyoro 0.2$.
The expected optical magnitude based on SEDs extended to optical region 
ranges from $\sim 17$ mag to $\sim 19$ mag (Hirashita et al. 1998).
Since the positional accuracy of the \iris~survey is quite good ($5''$ -- 
$10''$) and surface density of such bright galaxies is low enough,
these galaxies would be identified easily with optical counterparts.
On the contrary, the high-$z$ galaxies are expected to have fainter magnitudes 
($\sim 20$ mag) (Hirashita et al. 1998).
Thus the cross correlation with optical survey data such as POSS, or 
forthcoming SDSS would identify most of the low-$z$ IR-galaxies.
Evolutionary effect makes the mid/high-$z$ galaxy more abundant, and the 
fraction of such objects in the overlapped region will be higher.
Thus, color classification of galaxy redshift can be practicable using 
sets of bandpasses $50, \; 70$, and $150\; \mu$m with the aid of present and 
future optical surveys.

\section{SUMMARY AND CONCLUSIONS}

In this paper we examined the performance of the \iris~FIR all-sky survey, 
one of whose main purposes is to study galaxy evolution and formation.
We used a simple empirical model for galaxy number count estimation, 
by adopting a multicomponent SED model which consists of
cirrus and starburst components, and the nearby 
FIR luminosity function derived from that of {\sl IRAS} galaxies.

Our conclusions are as follows: 
\begin{enumerate}
\item A few $\times 10^6$ galaxies will be detected by the \iris~all-sky 
survey, which is 100 times larger number of detection than that of \iras.
\item Shorter wavelength bandpass ($50 - 70\; \mu$m) is suitable for 
intermediate-$z$ ($1 < z < 2.5$) galaxy detection, while longer wavelength
($120 - 150\; \mu$m) one is for high-$z$ ($2.5 < z$) and nearby ($z < 1$) 
galaxies.
\item We will detect the effect of galaxy evolution and can evaluate 
the amplitude of evolution at least in the nearby universe in the \iris~
survey, but still it is difficult to constrain which type of evolution 
takes place, from the number count alone.
On the contrary, the intensity of the CIRB complementarily provides us 
much tighter constraint on the form of galaxy evolution at very high-$z$.
\item Redshift estimation using galaxy FIR colors can be practicable using 
sets of bandpasses $50, \; 70$, and $150\; \mu$m with the aid of present and
future optical surveys.
\end{enumerate}

We first thank the anonymous referee for valuable suggestions and corrections
which improved the paper much.
We owe a great debt to Drs. Hideo Matsuhara, Takao Nakagawa, Mitsunobu 
Kawada, and other members of the  \iris~mission for useful discussions 
and comments.
We also acknowledge Drs. Chris P. Pearson, Gavin Dalton, Toru Yamada, 
Kimiaki Kawara for fruitful discussions.
We offer our gratitude to Profs. Mamoru Sait\={o}, Hiroki Kurokawa,
and Shin Mineshige for continuous encouragement.
Two of us (TTT and HH) acknowledge the Research Fellowships
of the Japan Society for the Promotion of Science for Young
Scientists.
This research has also made use of the NASA/IPAC Extragalactic Database (NED) 
which is operated by the Jet Propulsion Laboratory, California Institute of 
Technology, under contract with the National Aeronautics and Space 
Administration.
We also made extensive use of the NASA's Astrophysics Data System Abstract 
Service (ADS).

\begin{appendix}
\section{APPENDIX : OTHER FUTURE MISSIONS}

We show the expected number counts and redshift 
distributions of galaxies in other future missions now planned, 
{\sl SOFIA, SIRTF, FIRST}, and {\sl LMSA}.
The present model is valid only within the mid-IR to sub-mm wavelengths 
because of the SED used, we restrict the count predictions within this range.
We calculated the number counts and redshift distributions in the 
same way as we did in the main text.
In this appendix, we only consider the case of $q_0 = 0.1$ cosmology with
pure luminosity evolution ($Q = 1.4$).
Though some of the facilities are not planned to be used for large-area 
survey, we uniformly calculated the number counts per unit solid angle, 
and the redshift distributions in the whole sky.
We applied the detection limits from published papers or from the web pages
of the missions if available.

\subsection{\sl SOFIA}

Stratospheric Observatory for Infrared Astronomy ({\sl SOFIA}) is 
an airborne observatory with a 2.5-meter telescope installed in a 
Boeing 747 aircraft, which will be flying in 2001.
The detection limits ($1\sigma$ noise level) for 1 hour exposure 
are 7 mJy at 100 $\mu$m and 3.5 mJy at 450 $\mu$m (Becklin 1997).
Based on these values we calculated the number count and redshifts
with $5\sigma$-detection limits (obtained by simply multiplying factor 5 to 
the above limits).
The galaxy number counts are presented in the upper-left panel of Fig. 10, 
and the redshift distributions are shown in the upper-left panel of Fig. 11.

\subsection{\sl SIRTF}

The Space Infrared Telescope Facility ({\sl SIRTF}) is the 4th 
and final element in NASA's family of ``Great Observatories''.
{\sl SIRTF} consists of a 0.85-meter telescope and three cooled 
instruments, IRAC, MIPS, and IRS, capable of performing imaging 
and spectroscopy in the $3 - 180\;\mu$m wavelength range.
It is planned to be launched in December 2001.
We present the predictions for MIPS (Multiband Imaging Photometer 
for {\sl SIRTF}) at wavelengths 23.5, 70, and 160 $\mu$m.
The 5$\sigma$-detection limits are 370 $\mu$Jy, 1.4 mJy, and 7.5 mJy,
respectively, by 500-s exposure (Heim et al. 1998).
{\sl SIRTF} will survey 60 \% of the whole sky.
The upper-left panel of Fig. 10 depicts the number counts, and the 
upper-left panel of Fig. 11 shows the redshift distribution.

\subsection{\sl FIRST}

The Far Infrared and Submillimetre Telescope ({\sl FIRST}) is the 4th
cornerstone mission in the European Space Agency's Horizons 2000 programme,
implemented in collaboration with NASA.
{\sl FIRST} will perform photometry and spectroscopy in the $80 - 670\;\mu$m
range.
Three instruments have been provisionally selected: HIFI, PACS, and SPIRE.
Imaging photometry is performed by PACS ($80 - 210\;\mu$m) and SPIRE 
($200 - 670\;\mu$m).
The $5\sigma$-detection limits (by 1 hour exposure) for PACS and SPIRE are 
5 mJy and 3 mJy, respectively (Pilbratt 1998).
We show 90-$\mu$m number count for PACS and 250, 350, and 500 $\mu$m 
for SPIRE in the lower-left panel of Fig. 10.
Redshift distributions of the objects detedted at these bandpasses are
presented in the lower-left panel of Fig. 11.

\subsection{\sl LMSA}

The Large Millimeter and Submillimeter Array ({\sl LMSA}) is the
ground-based radio facility proposed in Japan. 
In the current design concept the array will consist of 50 10-m antennas 
and will be covering observing frequencies from 80 to 800 GHz. 
The array will be located at very high site to realize sub-arcsec 
resolution imaging at very high frequencies. 
We present the number counts expected by {\sl LMSA} at 350, 450, 650, and
850 $\mu$m.
The $5\sigma$ sensitivities at these wavelengths (yearly mean values)
are 3350, 1800, 750, and 115 $\mu$Jy/beam, respectively, by 8-hour 
integration (Kawabe \& Kohno 1998, private communication).
Under the best weather condition in winter, much higher sensitivities 
can be achieved.
The galaxy number counts are shown in the lower-right panel of Fig. 10.
The lower-right panel of Fig. 11 illustrates the expected redshifts.
In the lower-right panel of Fig. 10, we compared our prediction with the 
850 $\mu$m number counts recently obtained by {\sl SCUBA} in JCMT.
Filled red circle represents the value obtained by Smail et al. (1998), 
filled square shows the result in Hubble Deep Field (Hughes et al. 1998),
and filled triangle is the number count in Lockman Hole (Barger et al. 1998).
In spite that we include the evolution in this calculation, our prediction
is an order of magnitude smaller than those results.
Thus, a very strong evolution is suggested by sub-mm observations.
It requires further detailed analyses of the present observational data
to study the history of galaxy evolution in wide range of wavelength.

\end{appendix}

\newpage

\begin{center}
{\large \bf Figure Captions}
\end{center}

\figcaption{
The model spectral energy distributions (SEDs) of galaxies with $\lir = 
10^8 L_\odot - 10^{14} L_\odot$. 
The solid curves represent our model SEDs with various $\lir$ (the lowest one:
$\lir = 10^8 L_\odot$; level interval: $\Delta \log \lir = 1.0$).
We also present the observed SEDs of two representative IR galaxies, 
M82 (Klein, Wielebinski, \& Morsi 1988), IRAS F$10214+4724$ 
(Rowan-Robinson et al. 1993), and SMM$02399-0136$ (Ivison et al. 1998).
Open triangles stand for the observed SED of M82,
open squares depict the SED of IRAS F$10214+4724$, and 
open diamonds represent the SED of SMM$02399-0136$.
Small downward arrows show upper-limit values.
The IR luminosity of M82 is $\sim 4 \times 10^{10} L_\odot$ and that of 
IRAS F$10214+4724$ is $\sim 3 \times 10^{14} L_\odot$, but the latter has 
been revealed to be magnified by gravitational lensing by a factor of 
$\sim 20$ (Broadhurst \& Leh\'{a}r 1995).
Its $\lir$ turns out to be $\sim 10^{13} L_\odot$, and its SED is 
corrected in this figure.
SMM$02399-0136$ is also a lensed object, and its unlensed $\lir$ is 
$\sim 4 \times 10^{13}$ (Ivison et al. 1998).
The SED of this galaxy is also corrected for the lensing.
Their SEDs are approximately fitted by our models with corresponding $\lir$.
}

\figcaption{
The local infrared luminosity function (LF) of galaxies $\phi_0 (\lir)$.
This is constructed from the LF of Soifer et al. (1987b).
Since their LF is defined by $L_{60\, \mu {\rm m}}$, we converted it into 
$\lir$ (the IR luminosity in the wavelength range of $3\;\mu{\rm m} - 1 \; 
{\rm mm}$), using our model SEDs.
The filled squares represent the converted data points, and solid line
stands for the fitted analytical function presented in the main text
(see eq. \ref{lf}).
Dotted line is shown for the $\lir$~ of the `knee' of our LF.
}

\figcaption{
The differential flux-number
relation for \iras~60-$\mu$m databases and our model predictions.
Open squares represent the data points for \iras~PSC, and filled squares,
HH87.
The no-evolution model tracks are also shown by black solid curves.
Upper panel shows the case of $q_0 = 0.1$, and lower panel, of $q_0 = 0.5$.
We performed least-square fitting of our evolutionary models (see section 
\ref{evpar}) on these data points.
When we adopt all data points with equal weight, the most likely value 
for the parameters are $(P\,,\; Q) = (1.3\,,\; 0.7) \; (q_0 = 0.1)$ and 
$(1.4\,, \; 0.8) \; (q_0 = 0.5)$.
We obtain higher values $(P\,,\; Q) = (2.7\,,\; 1.4) \; (q_0 = 0.1)$ and 
$(2.8\,, \; 1.5) \; (q_0 = 0.5)$ when the four statistically poorest points
around $\log S \;[{\rm Jy}] \sim -0.5$ are removed.
We adopt the latter in our studies.
}

\figcaption{
The integrated galaxy number count $(q_0 = 0.1 \; {\rm and} \; 0.5)$ at 
60-$\mu$m and 175-$\mu$m.
The results of QMW \iras~galaxy survey (Rowan-Robinson et al. 1991) and 
the \iso~Lockman Hole survey (Kawara et al. 1998) are also presented.
The upper panels represents the 60-$\mu$m prediction.
The agreement of the model and the QMW count is excellent.
The lower panels shows the 175-$\mu$m number count.
The horizontal errors are caused by the uncertainty of flux calibration.
Within the error, the model agrees with the \iso~result.
}

\figcaption{
The expected cosmic infrared background (CIRB) spectra with $q_0 = 0.1$ 
and $q_0 = 0.5$.
The background intensity depends on $q_0$, so that larger $q_0$ leads 
to smaller $I(\nu)$.
We also put the observational constraints on CIRB obtained by \cobe~
measurement.
Open red triangles represent the ``dark sky'' (see section \ref{coirb}) 
upper limit to the CIRB measured by \cobe~Diffuse Infrared Background 
Experiment (DIRBE).
Filled squares are the result of a similar analysis of the \cobe~Far 
Infrared Absolute Spectrophotometer (\firas) high-frequency data, 
after removal of the cosmic microwave background signal.
Open blue triangles represent the residual signal of \dirbe~after 
removing the contributions from the model foreground sources such as 
Galactic diffuse emission, and interplanetary dust emission.
The green horizontal lines represent the upper limit derived from 
Mather et al. (1994), and solid orange broken lines represent the 
approximate value of the residual light reported by Puget et al. (1996), 
Fixsen et al. (1998) and Hauser et al. (1998) with errors.
The CIRB constraint on the magnitude of galaxy evolution is tighter 
than that obtained from the number count.
}

\figcaption{
The galaxy number count predictions at assumed \iris~four 
bandpasses.
The upper four panels depict the number counts with $q_0 = 0.1$ and the 
lower four, $q_0 = 0.5$; 
from left to right, $\lambda = 50, 70, 120,$ and $150\,\mu$m.
The black solid lines show the no-evolution prediction.
The red lines and cyan lines represent the case of pure luminosity 
evolution and pure density evolution, respectively.
The green dot-dashed lines denote the \iris~FIS flux detection 
limit at each waveband.
}

\figcaption{
The expected redshift distribution of galaxies 
detected in the assumed four bands.
Figure 7a illustrates the prediction at 50-$\mu$m band.
Figures 7b, 7c, and 7d are the same as Fig. 7a, except that they show 
the predictions at 70-, 120-, and 150-$\mu$m bands, respectively.
Solid curves denote the predictions for $q_0 = 0.1$, and dashed curves, 
$q_0 = 0.5$.
Larger $q_0$ leads to deeper $z$-distribution.
The steep increases of galaxy number at $z < 0.5$ seen in 
Figs. 7c and 7d are attributed to the peak at $\sim 100\; \mu$m, which 
originates from big grains, in the SED of low-$\lir$ galaxies. 
}

\figcaption{
An example of the redshift distribution of detected galaxies with various 
$\lir$ ($q_0 = 0.1$, no evolution). 
Solid curves represent the galaxies with low-$\lir$, 
dashed curves the intermediate-$\lir$, and dot-dashed curves 
the high-$\lir$. }

\figcaption{
The loci of galaxies with various $\lir$ on the FIR 
color--color ($\log S_{50}/S_{70} - \log S_{50}/S_{150}$) plane.
Filled squares are put on every $\Delta z = 0.5$ interval.
Red filled squares represent the colors of the low-$z$ ($z < 1$) galaxies, 
green filled squares the intermediate-$z$ ($1 < z < 2.5$), and 
blue filled squares the high-$z$ ($2.5 < z$) ones.
For the sources fade away from the \iris~detection limits, we do not
show their loci and symbols.
To see how the starburst changes the color of cool cirrus 
galaxies, we put the galaxies of $\lir = 10^{10.1}\,L_\odot - 10^{10.9}\,
L_\odot$ with an interval of $\Delta \log \lir = 0.1$ (denoted by red open 
squares from the top-right to down-left).
The domain which is occupied mainly by low-$z$, low-$\lir$ galaxies are shown 
roughly by red dotted rhomboid.
Green rectangle shows the domain of intermediate-$z$ ones, and blue
rectangle shows the region where high-$z$ galaxies to be located.
}

\figcaption{
Integrated number counts of galaxies expected for {\sl SOFIA}, {\sl SIRTF}, 
{\sl FIRST}, {\sl LMSA}.
{\sl SOFIA} (upper-left) : 
Blue and red solid lines represent our 120 $\mu$m and 450 $\mu$m 
predictions, respectively.
The vertical blue and red dot-dashed lines show the $5\sigma$-detection 
limit of {\sl SOFIA} with 1-hour exposure.
We adopted $q_0 = 0.1$ with pure luminosity evolution ($Q=1.4$) in this and
the following figures.
{\sl SIRTF} (upper-right) : 
Integrated number counts of galaxies expected for {\sl SIRTF}.
Blue, green, and red solid lines represent our 25, 70, and 160 $\mu$m 
predictions, respectively.
The vertical blue, green, and red dot-dashed lines show the 
$5\sigma$-detection limit of {\sl SIRTF} MIPS with 500-s exposure.
{\sl FIRST} (lower-left) : 
Integrated number counts of galaxies expected for {\sl FIRST}.
Blue, green, yellow and red solid lines represent our 90, 250, 350, 
and 500 $\mu$m predictions, respectively.
The vertical blue, green, yellow and red dot-dashed lines show the 
$5\sigma$-detection limit of {\sl FIRST} PACS ($90\;\mu$m) and SPIRE 
(250, 350, and 500 $\mu$m) with 1-hour exposure.
{\sl LMSA} (lower-right) : 
Integrated number counts of galaxies expected for {\sl LMSA}.
Blue, green, yellow and red solid lines represent our 350, 450, 650, 
and 800 $\mu$m predictions, respectively.
The vertical blue, green, yellow and red dot-dashed lines show the 
$5\sigma$-detection limit of {\sl LMSA} with 8-hour exposure.
Filled red circle represents the galaxy number count reported 
by Smail et al. (1998), filled square shows the result in Hubble 
Deep Field (Hughes et al. 1998), and filled triangle is the count 
in Lockman Hole (Barger et al. 1998), all obtained with {\sl SCUBA} 
in JCMT at 850 $\mu$m .
}

\figcaption{
The redshift distribution of galaxies detected with {\sl SOFIA}, 
{\sl SIRTF} MIPS, {\sl FIRST}, and {\sl LMSA}
assuming the whole-sky observation;
upper-left : {\sl SOFIA}, 
upper-right : {\sl SIRTF}, 
lower-left : {\sl FIRST}, 
and lower-right : {\sl LMSA}.
}

\newpage

\begin{center}
Table 1 : Adopted Evolutionary Parameters
\vspace*{5mm}

\begin{tabular}{ccc} \hline \hline
$q_0$ & 0.1 & 0.5 \\ \hline
$P$ & 2.7 & 2.8 \\
$Q$ & 1.4 & 1.5 \\ \hline
\end{tabular}
\end{center}

\newpage

\begin{center}
Table 2 : Adopted Flux Limits of \iris~FIS
\vspace*{5mm}

\begin{tabular}{cc} \hline \hline
Wavelength & $5 \sigma$-detection limit \\

[$\mu $m] & [mJy] \\ \hline
50 & 20 \\
70 & 15 \\
120 & 30 \\
150 & 50 \\ \hline
\end{tabular}
\end{center}

\newpage

\begin{figure}
\begin{center}
    \includegraphics[width=14cm,clip]{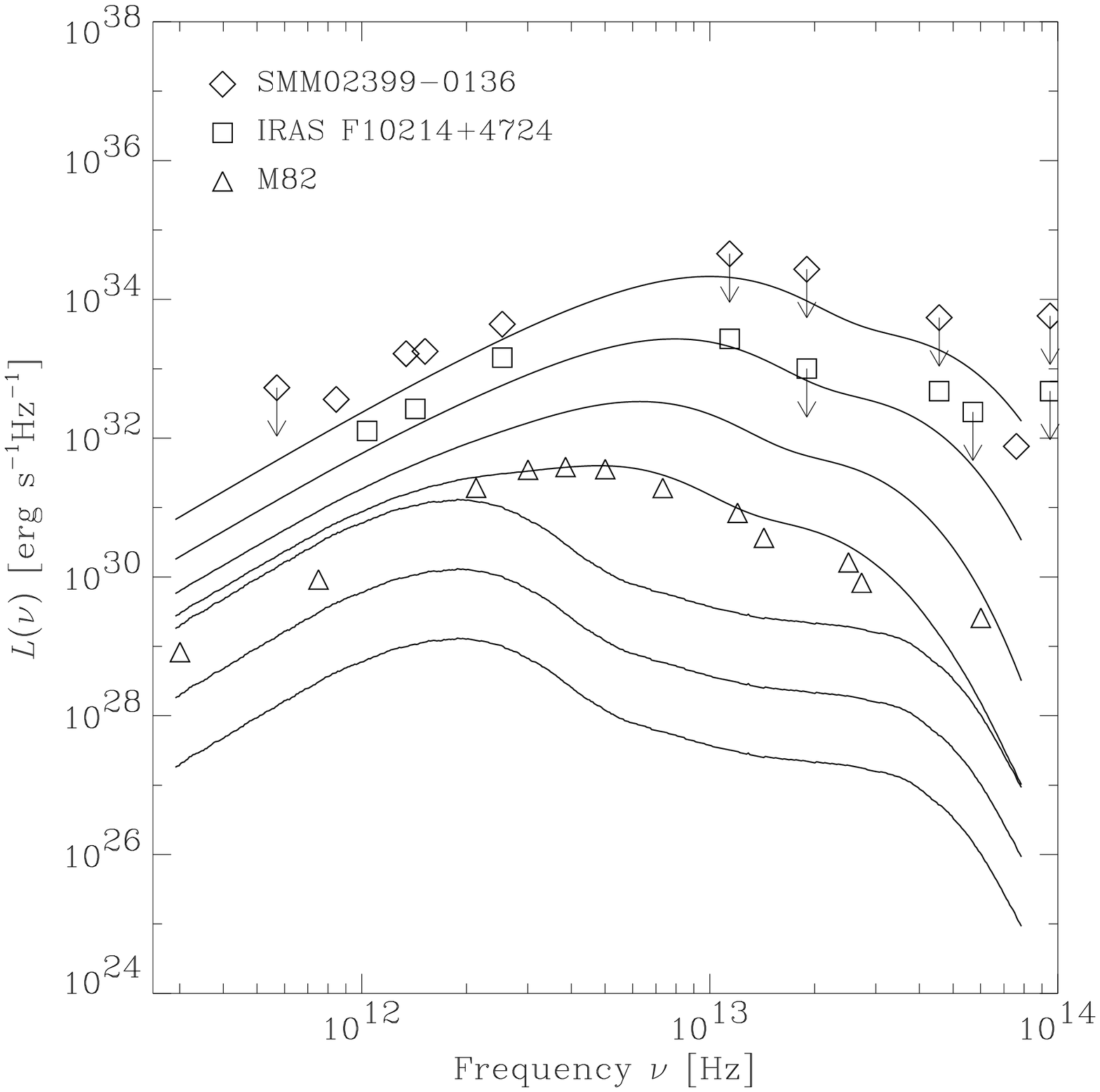}

Figure 1
\end{center}
\end{figure}

\newpage

\begin{figure}
\begin{center}
    \includegraphics[width=14cm,clip]{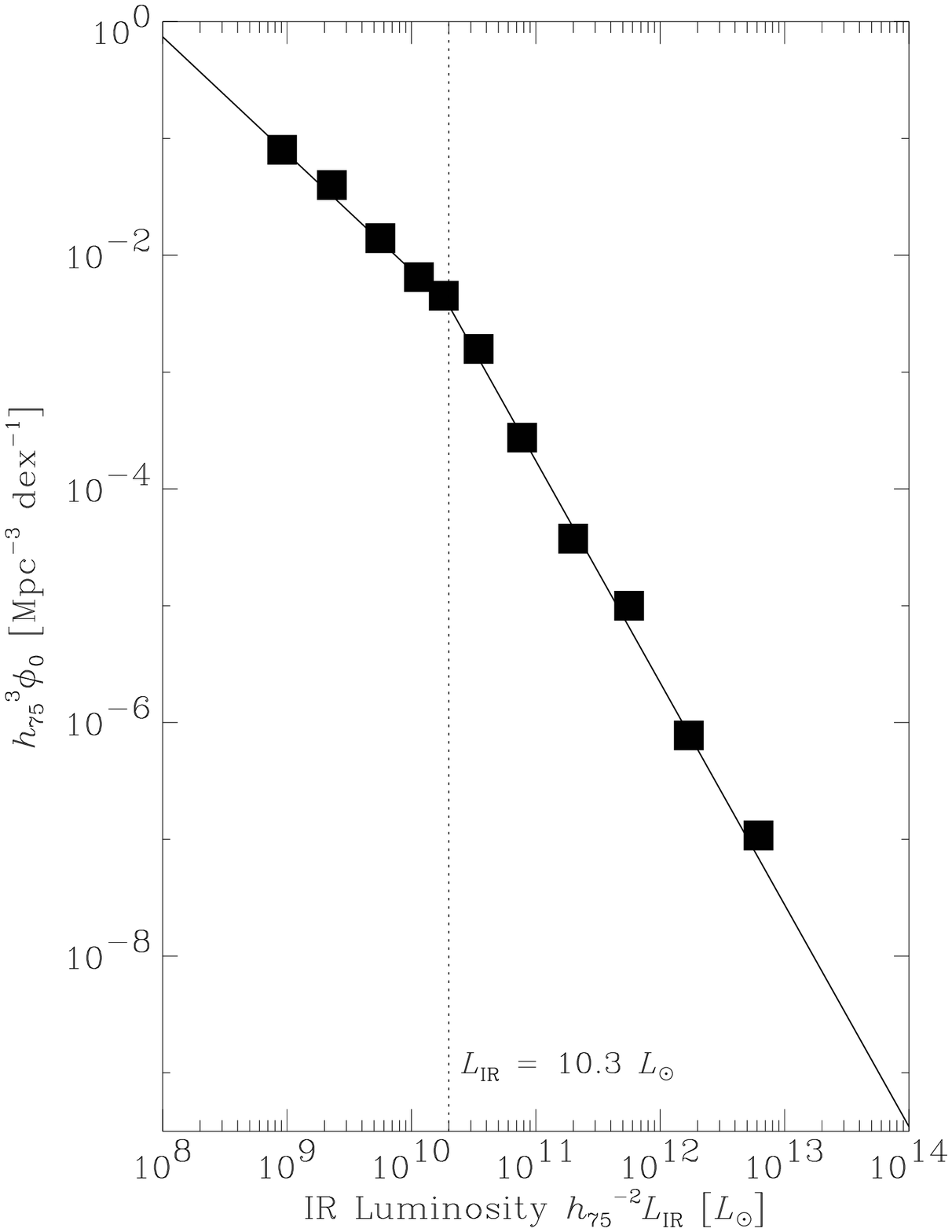}

Figure 2
\end{center}
\end{figure}

\newpage

\begin{figure}
\begin{center}
    \includegraphics[width=14cm,clip]{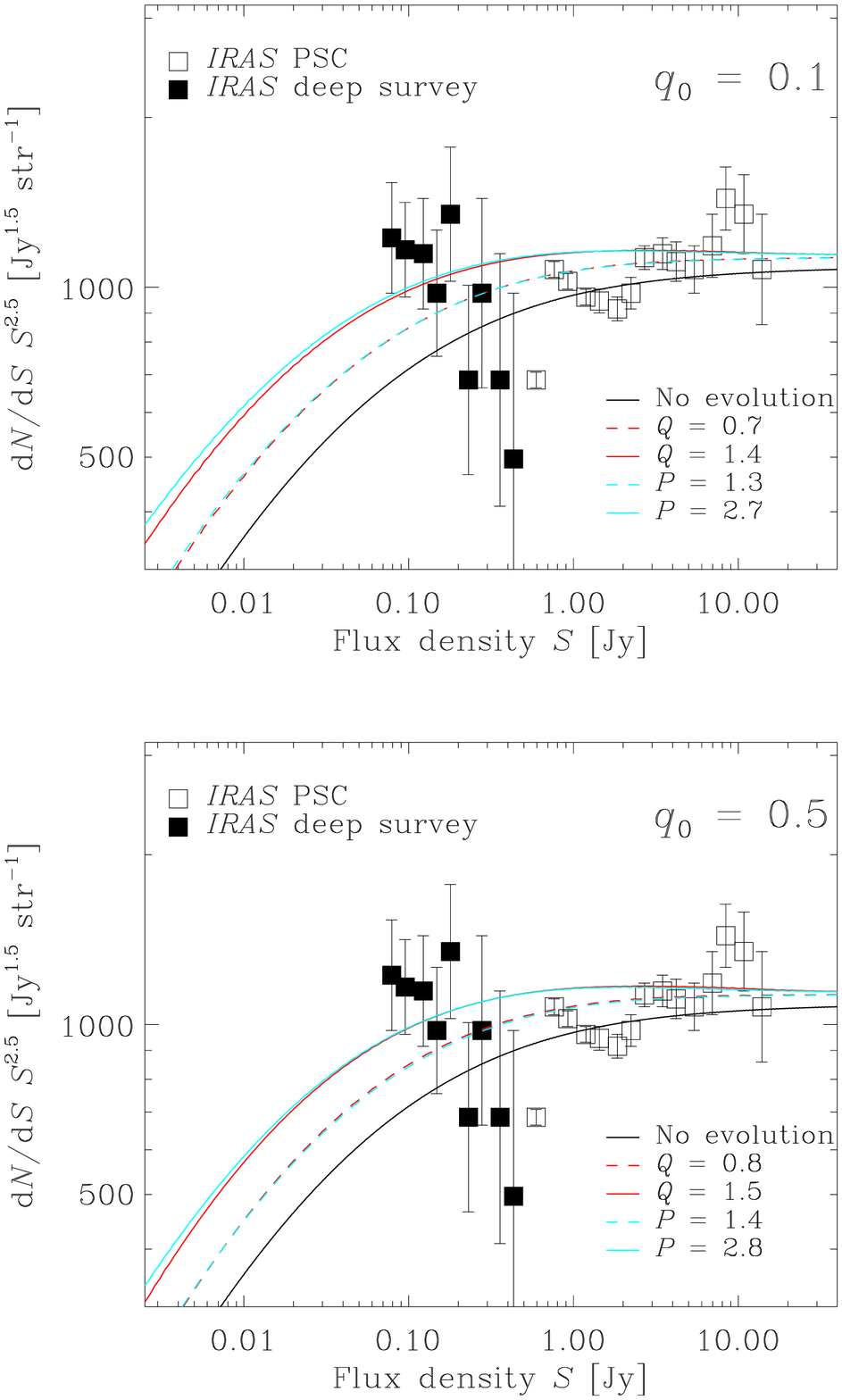}

Figure 3
\end{center}
\end{figure}

\newpage

\begin{figure}
\begin{center}
    \includegraphics[width=14cm,clip]{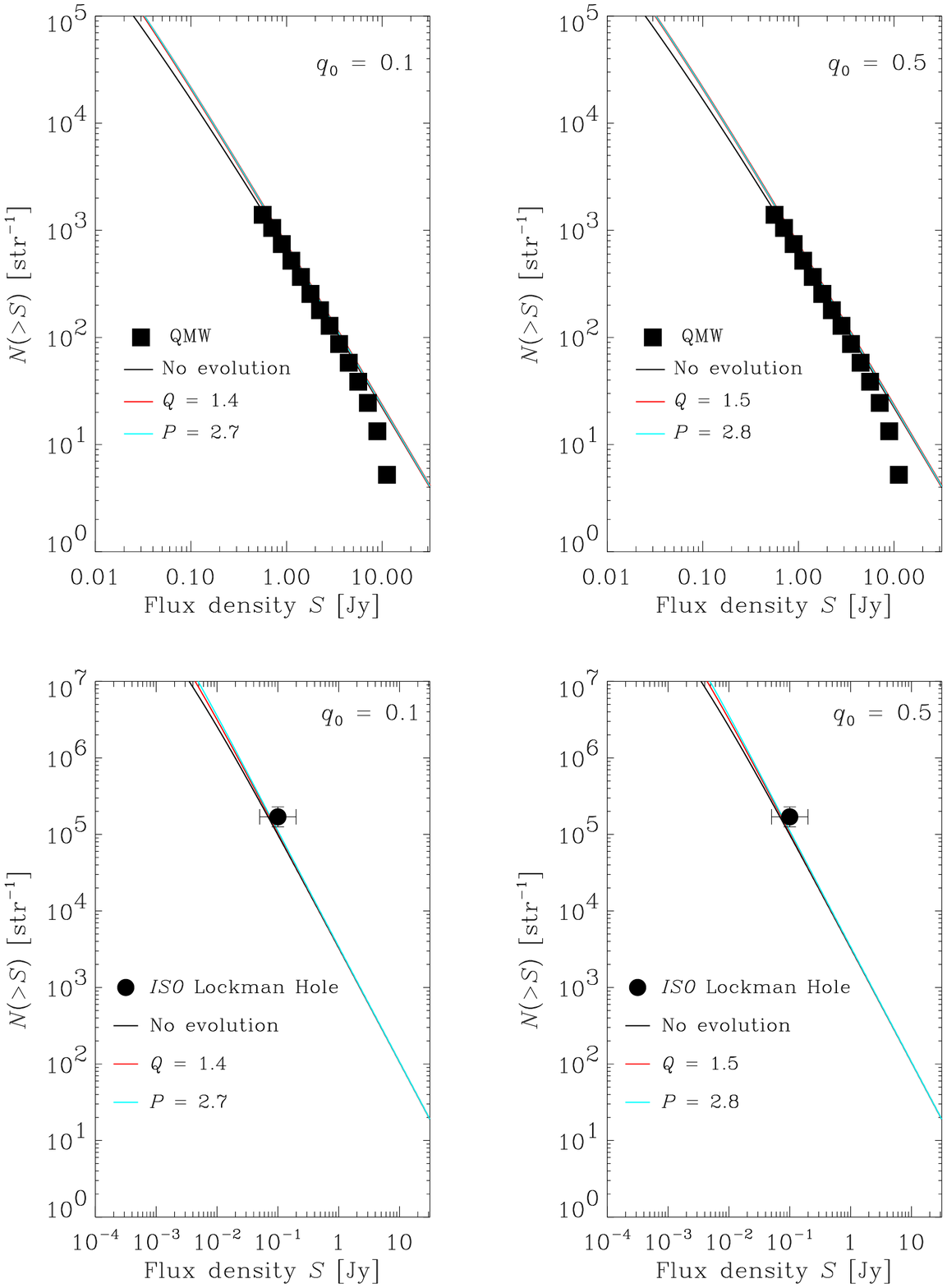}

Figure 4
\end{center}
\end{figure}

\newpage

\begin{figure}
\begin{center}
    \includegraphics[width=14cm,clip]{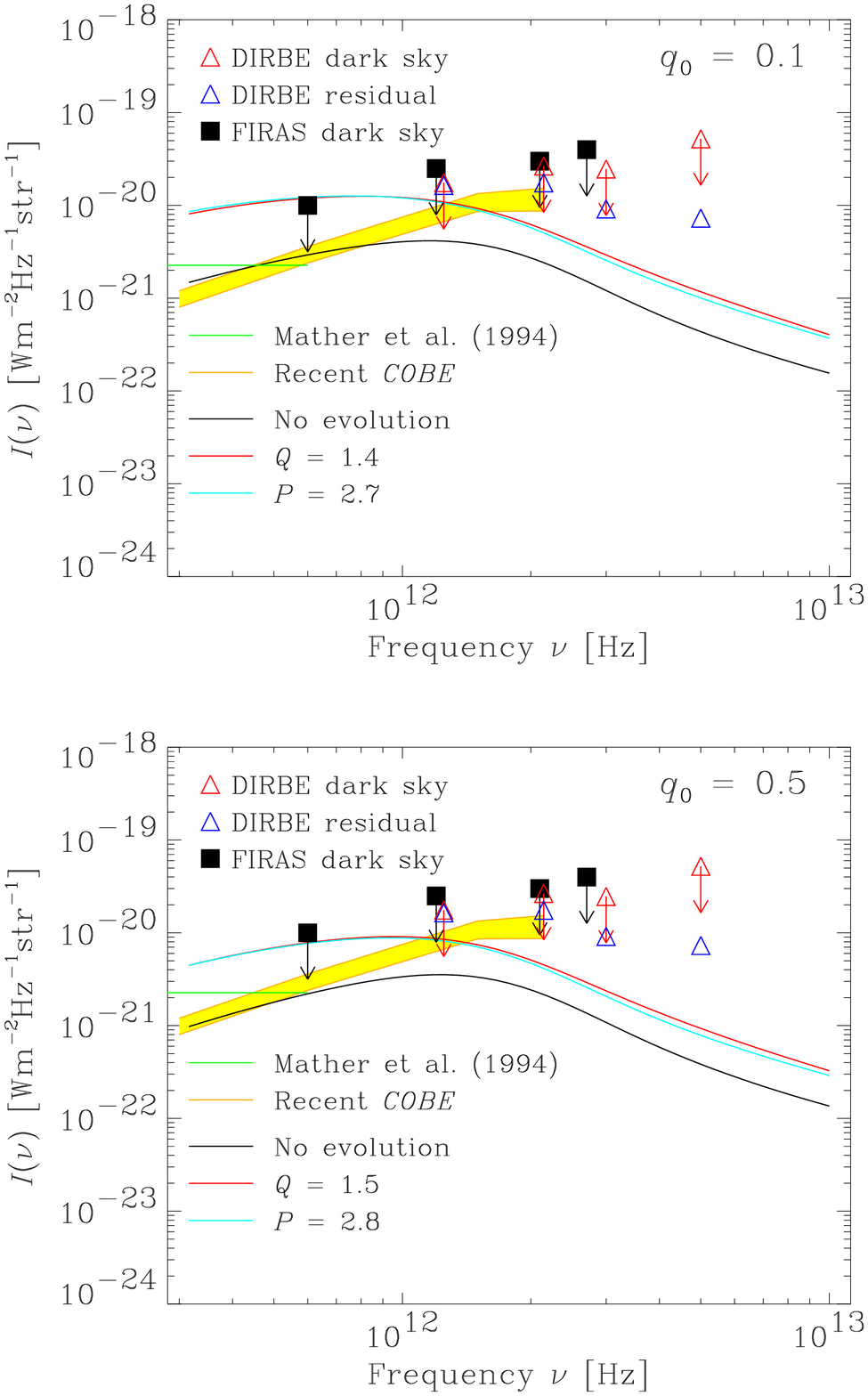}

Figure 5
\end{center}
\end{figure}

\newpage

\begin{figure}
\begin{center}
    \includegraphics[angle=180,width=14cm,clip]{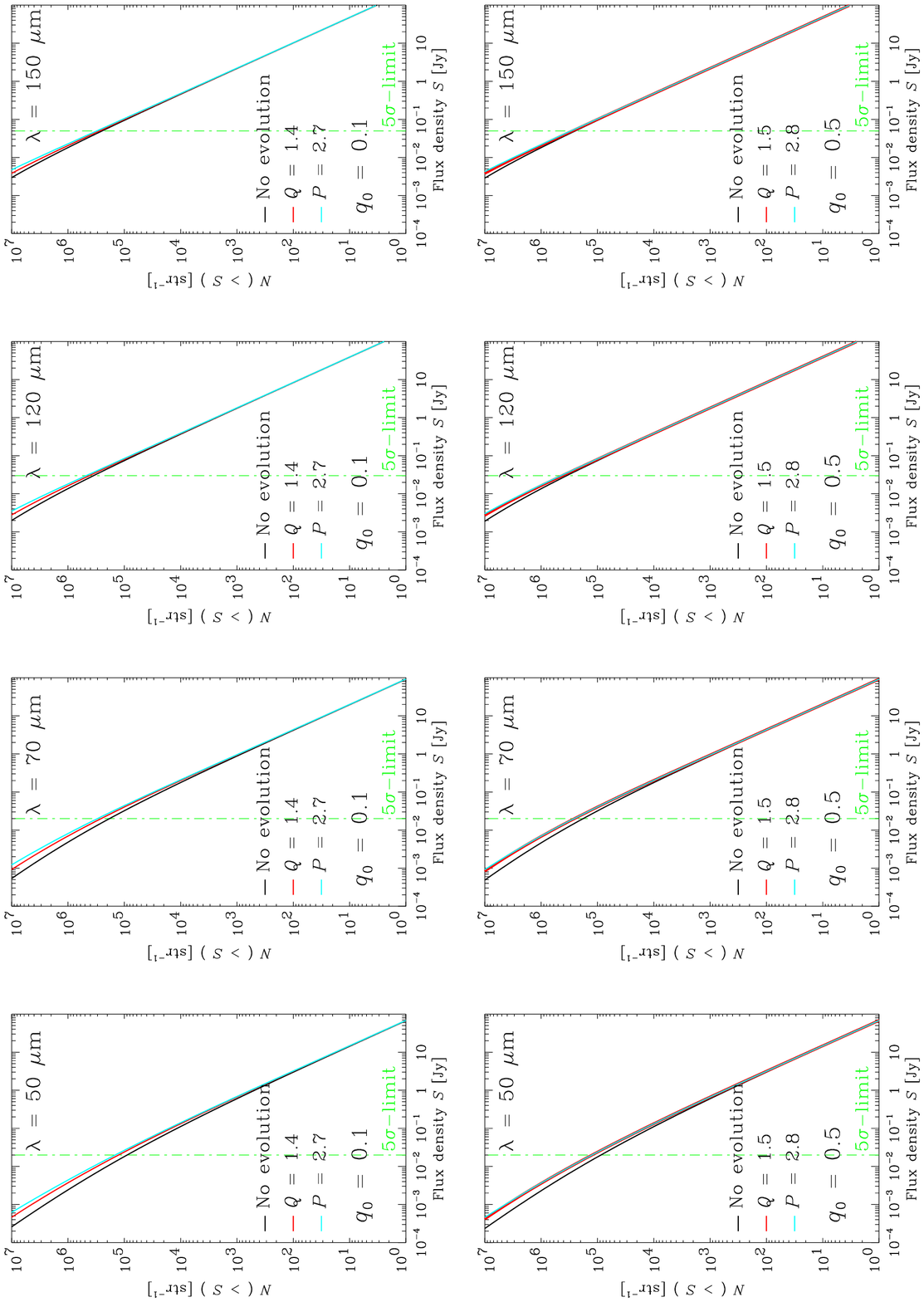}

Figure 6
\end{center}
\end{figure}

\newpage

\begin{figure}
\begin{center}
    \includegraphics[angle=180,width=14cm,clip]{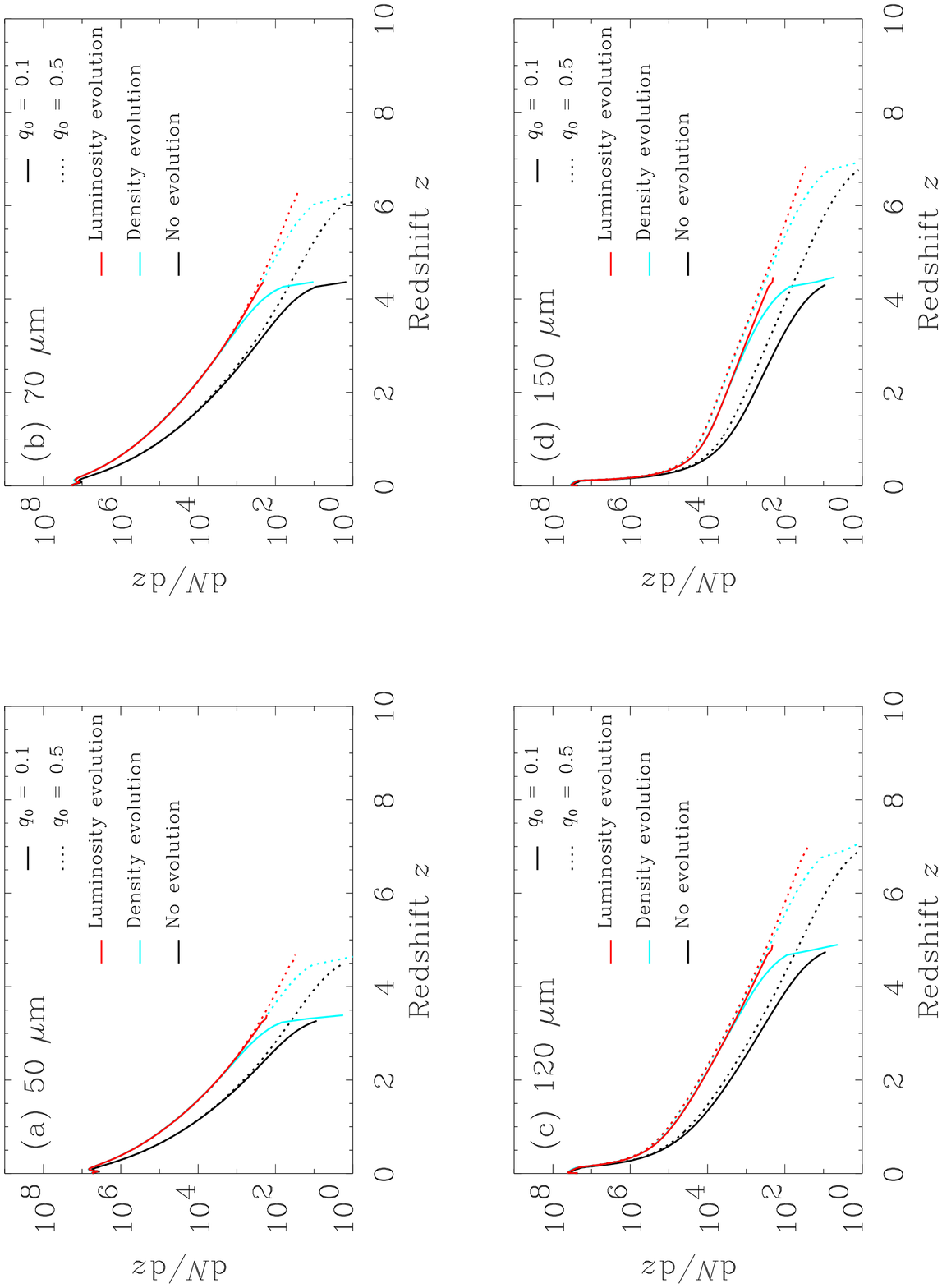}

Figure 7
\end{center}
\end{figure}

\newpage

\begin{figure}
\begin{center}
    \includegraphics[width=14cm,clip]{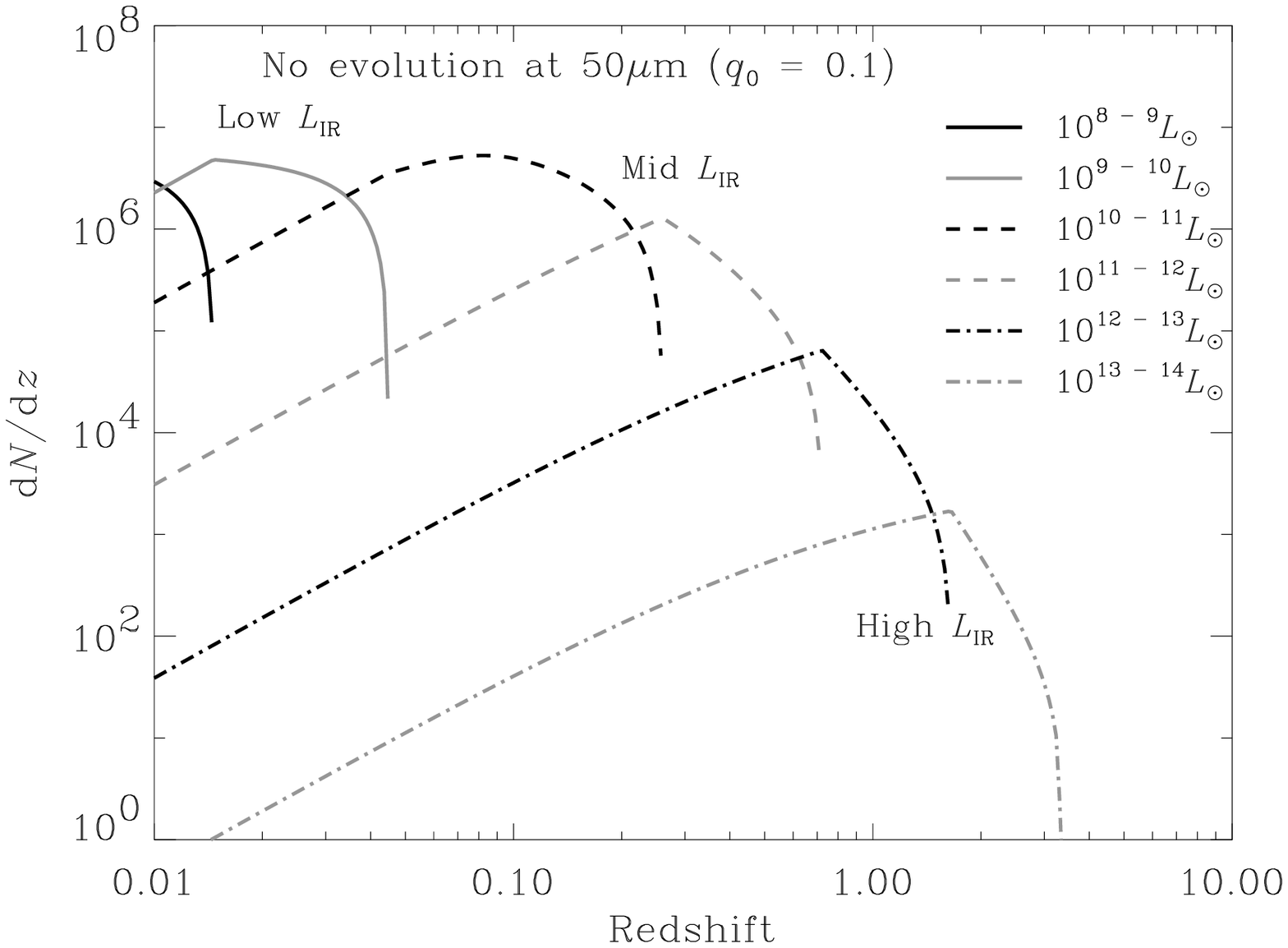}

Figure 8
\end{center}
\end{figure}

\newpage
\begin{figure}
\begin{center}
    \includegraphics[width=14cm,clip]{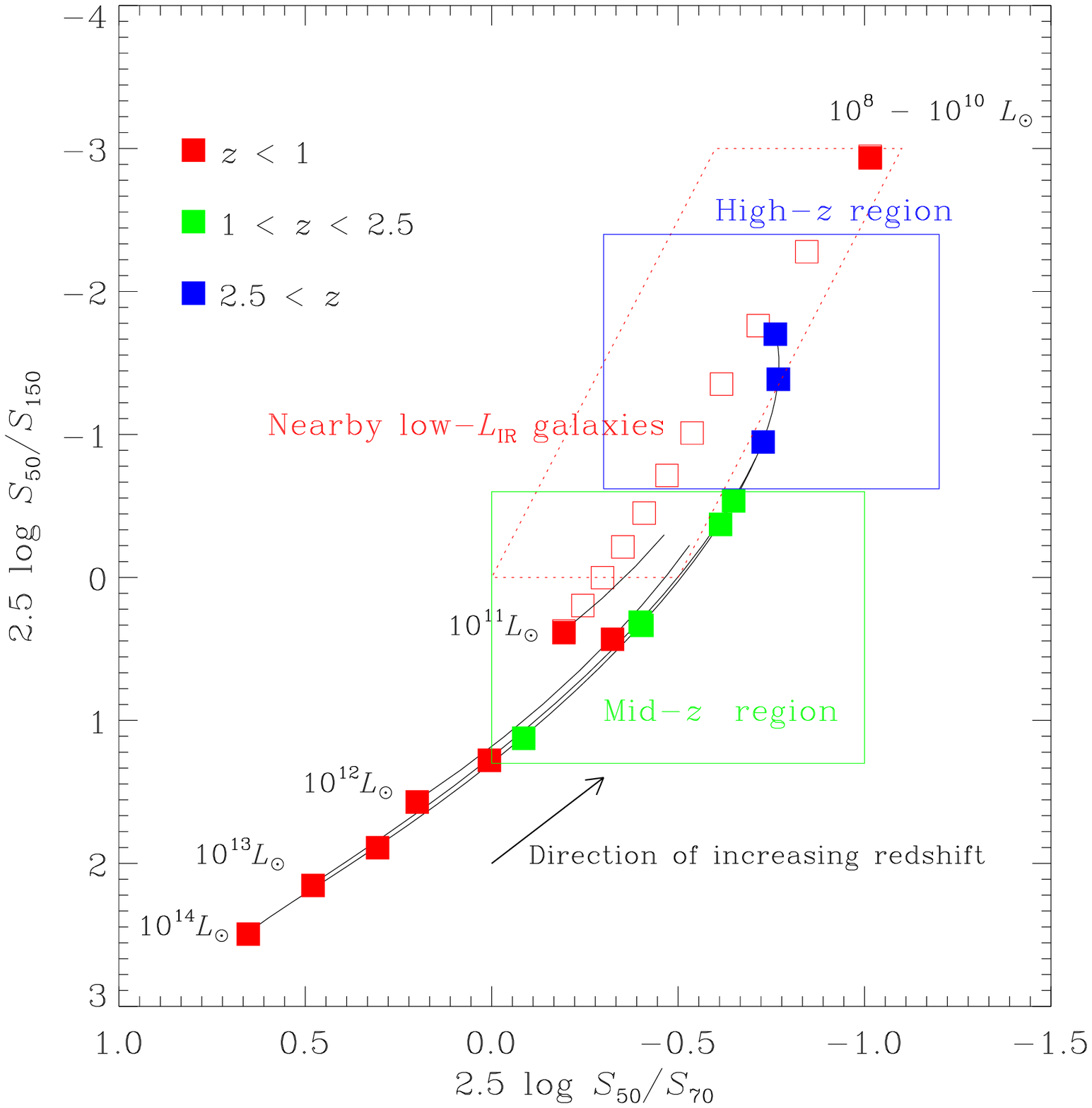}

Figure 9
\end{center}
\end{figure}

\newpage

\begin{figure}
\begin{center}
    \includegraphics[width=14cm,clip]{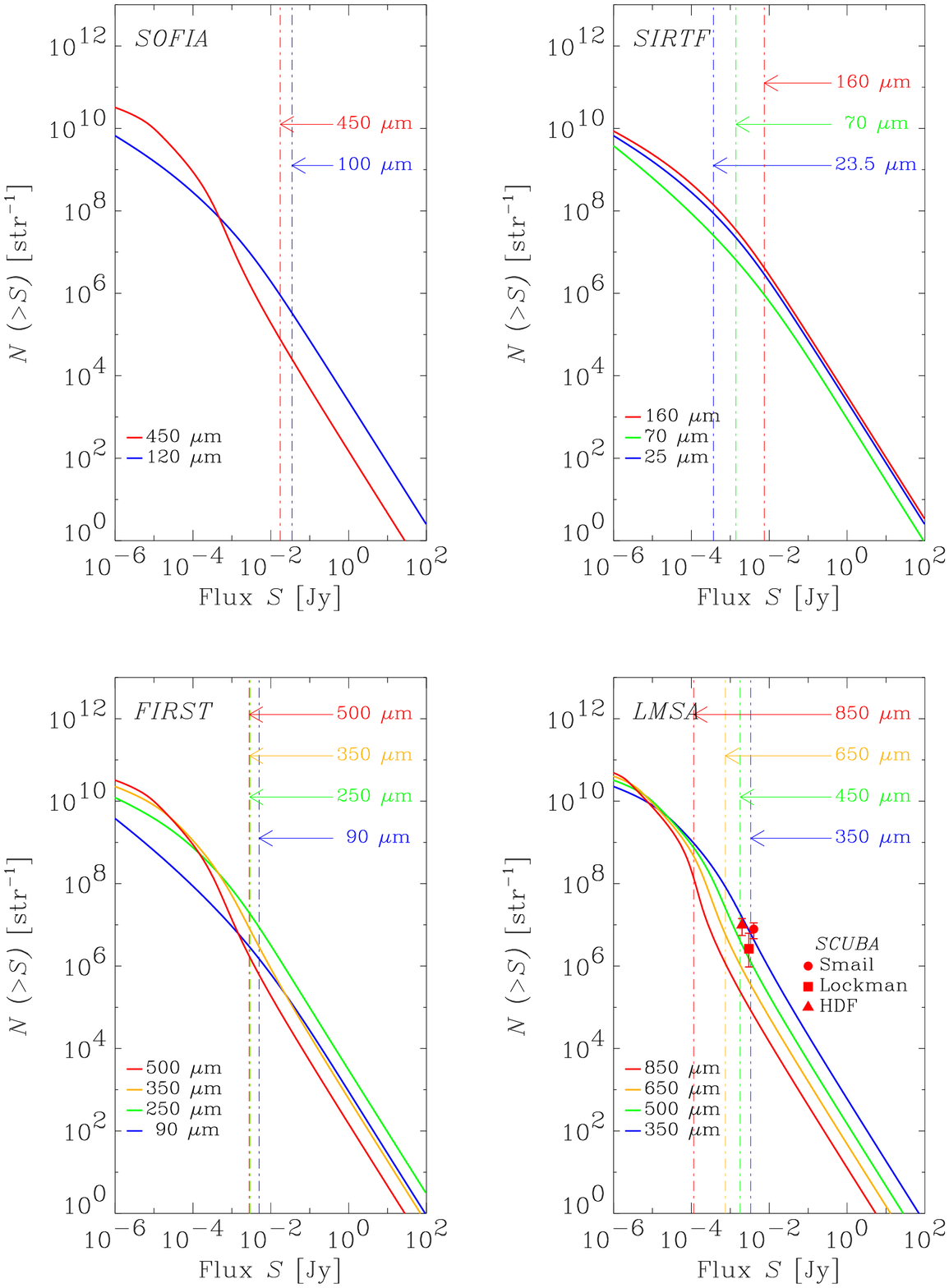}

Figure 10
\end{center}
\end{figure}

\newpage

\begin{figure}
\begin{center}
    \includegraphics[angle=180,width=14cm,clip]{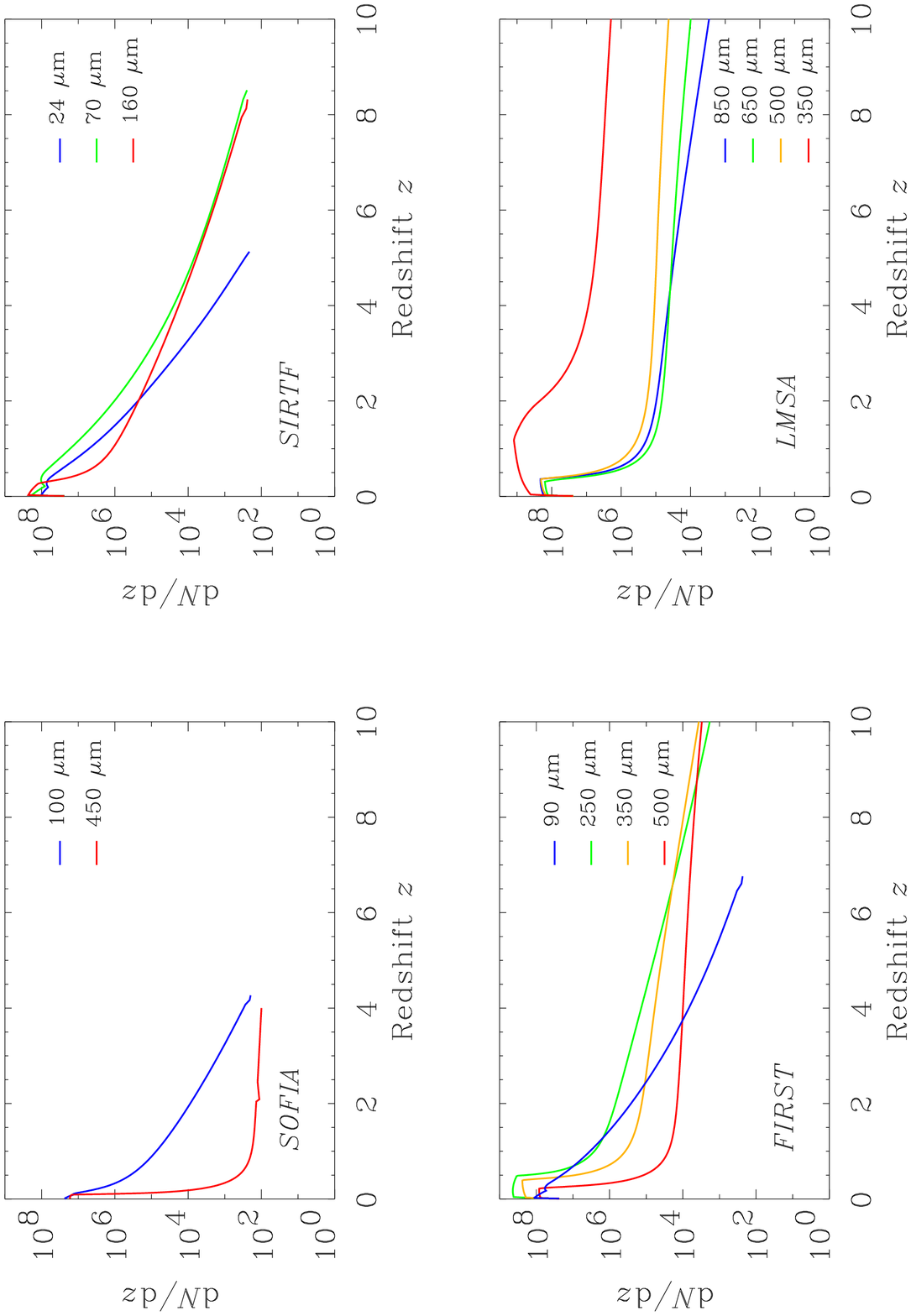}

Figure 11
\end{center}
\end{figure}

\end{document}